\documentclass[10pt,journal,compsoc]{IEEEtran}
\ifCLASSOPTIONcompsoc
  \usepackage[nocompress]{cite}
\else
  \usepackage{cite}
\fi
\ifCLASSINFOpdf
\else
\fi

\usepackage{amsmath,graphicx}
\usepackage{amsfonts}
\usepackage{amssymb}
\usepackage{mathptmx}
\usepackage{subfigure}
\usepackage{multirow}
\usepackage{bbm}
\usepackage{amsthm}
\usepackage{tikz}
\usepackage{hyperref}
\usepackage{booktabs}
\usepackage{enumitem}

\newtheorem{definition}{Definition}[section]

\begin{document}
\title{Attribute Inference Attack of Speech \\ Emotion Recognition on Federated Learning}

\author{Tiantian Feng,~\IEEEmembership{~Student Member, ~IEEE}
        Hanieh Hashemi,~\IEEEmembership{}
        Rajat Hebbar,~\IEEEmembership{}
        % Rahul Gupta,~\IEEEmembership{~Member, ~IEEE}
        Murali Annavaram,~\IEEEmembership{~Fellow,~IEEE},
        and~Shrikanth S. Narayanan,~\IEEEmembership{~Fellow,~IEEE}
        
\IEEEcompsocitemizethanks{\IEEEcompsocthanksitem Tiantian Feng is with the Department of Computer Science, USC, Los Angeles, CA, 90007. Hanieh Hashemi, Rajat Hebbar, Murali Annavaram, and Shrikanth S. Narayanan are with the Department of Electrical Engineering, USC, Los Angeles,
CA, 90007. email: tiantiaf@usc.edu.

% \IEEEcompsocthanksitem Rahul Gupta is with Amazon.com. email: gupra@amazon.com. \protect\\
}% <-this % stops a space

\thanks{}}


\markboth{}%
{Shell \MakeLowercase{\textit{et al.}}: Bare Advanced Demo of IEEEtran.cls for IEEE Computer Society Journals}



\IEEEtitleabstractindextext{
\begin{abstract}

SER systems often acquire and transmit speech data collected at the client-side to remote cloud platforms for further processing. However, speech data carry rich information about emotions and other sensitive demographic traits like gender and age. Given the increasing emphasis on privacy considerations, it is desirable for SER systems to classify emotion constructs while preventing inferences of other sensitive information embedded in the speech signal. Federated learning (FL) is a distributed machine learning algorithm that coordinates clients to train a model collaboratively without sharing local data. While FL improves privacy by preventing direct sharing of client’s data, recent works have demonstrated that FL is still vulnerable to many privacy attacks. To assess the information leakage of SER systems trained using FL, we propose an attribute inference attack framework that infers sensitive attribute information of the clients from shared model updates. As a use case, we empirically evaluate our approach for predicting the client's gender using three SER benchmark datasets: IEMOCAP, CREMA-D, and MSP-Improv. We show that the attribute inference attack is achievable for SER systems trained using FL. We perform further analysis and make observations such as the most information leakage comes from the first layer in the SER model.

\end{abstract}

}

\maketitle

% is not selected <OR> if conference mode is selected - because compsoc
% conference papers position the abstract like regular (non-compsoc)
\IEEEdisplaynontitleabstractindextext

% \fi
\IEEEpeerreviewmaketitle

\ifCLASSOPTIONcompsoc
\IEEEraisesectionheading{\section{Introduction}\label{sec:introduction}}
\else

\section{Introduction}
\label{sec:introduction}
\fi

\IEEEPARstart{S}{peech} emotion recognition (SER) aims to identify emotional states conveyed in vocal expressions. Speech emotion recognition systems are currently deployed in a wide range of applications such as in smart virtual assistants \cite{lee2020study}, clinical diagnoses \cite{ramakrishnan2013speech,Bone2017SignalProcessingandMachine}, and education \cite{li2007speech}. A typical centralized SER system has three parts: data acquisition, data transfer, and emotion classification \cite{koolagudi2012emotion}. Under this framework, the client typically shares the raw speech samples or the acoustic features derived from the speech samples (to obfuscate the actual content of the conversation) to the remote cloud servers for emotion recognition. However, the same speech signal carries rich information about individual traits (e.g., age, gender) and states (e.g., health status), many of which can be deemed sensitive from an application point of view. Attribute inference attacks would aim to reveal an individual's sensitive attributes (e.g., age and gender) that they did not intend or expect to share \cite{gong2018attribute, mireshghallah2020privacy}. These undesired/unauthorized usages of data may occur when the service provider is not trustworthy (insider attack) or an intruder attacks the cloud system (outsider attack)~\cite{domingo2019privacy, kumar2018exploring, tabrizchi2020survey}.

Federated learning (FL) is a popular privacy-preserving distributed learning approach that allows clients to train a model collaboratively without sharing their local data  \cite{mcmahan2017communication}. In an FL setting, during the training process, a central server aggregates model updates from multiple clients. Each client generates such model updates by locally training a model on the private data available at the client. This machine learning approach reduces information leaks compared to classical centralized machine learning frameworks since personal data does not leave the client. Therefore, this distributed learning paradigm can be a natural choice for developing real-world multiuser SER applications as sharing raw speech or speech features from users' devices is vulnerable to attribute inference attacks.

\textbf{Attacks in Federated Learning: } Arguably, while sharing model updates can be considered more privacy preserving than sharing raw data, recent works have demonstrated that FL can be susceptible to a variety of privacy attacks, including membership inference attacks \cite{melis2019exploiting} and reconstruction attacks \cite{zhu2020deep, geng2021towards}. For instance, recent work has shown that the attacker can efficiently reconstruct a training image from the gradients \cite{zhu2020deep}. More recent works increasingly show that image reconstruction is also achievable through the model parameter updates even without accessing to the raw gradients \cite{geng2021towards}. On the other hand, prior work has demonstrated that the attacker can perform membership attacks in FL settings to infer whether a particular model update belongs to the private training data of a single participant (if the update is of a single participant) or of several participants (if the update is the aggregate) \cite{melis2019exploiting}. While existing research has demonstrated the vulnerability of FL training to privacy attacks in the CV domain, it is reasonable to believe that the shared model updates in training the SER model using the FL technique also introduce information leakage.

\textbf{Threat Model:} This work presents a detailed analysis of the attribute inference attack on the SER application trained in an FL setting. In general there are two sub-types of attacks based on what attacker can observe. In the black-box attack, the attacker can only observe the outputs of the model ($F(\mathbf x;\mathbf W)$) for any given input~\cite{shokri2017membership}. However, in the white-box attack, attacker can access the model parameters, intermediate values, and other model information as well~\cite{nasr2019comprehensive}. In this work, we support white-box attack in which the attacker knows all model parameters and hyper-parameters in the FL process, including learning rate, local epochs, local batch size, local sample size, and model architecture. The white-box attack is a realistic scenario in this setting because this information can be available to the attacker from any participating client or if the attacker operate as client itself. Any adversary that has access to the shared model updates can execute the attack. The attacker's goal is to infer sensitive attributes of the client using shared model updates (parameters/gradients) of SER applications trained under FL architecture. In this work, we consider gender prediction as the exemplary attribute inference attack task. We show that the adversary can effectively infer a client's gender attribute while training the SER model in an FL setup; we use the IEMOCAP \cite{busso2008iemocap}, Crema-D \cite{cao2014crema}, and MSP-Improv \cite{busso2016msp} datasets for the experiments. To the best of our knowledge, this is the first work to demonstrate that shared model updates that are communicated in FL to train an SER model can cause attribute information leakage (e.g., gender).

\section{SER Experimental Data Sets}
\label{sec:data}

In this work, we use three data sets for developing SER models and threat models. Due to the data imbalance issue in the IEMOCAP corpus, previous works use the four most frequently occurring emotion labels (neutral, sad, happiness, and anger) for training the SER model \cite{zhang2018attention}. In addition to this, we pick these four emotion classes because all three corpora contain these labels. \autoref{tab:dataset} shows the label distribution of utterances in these corpora. The details of these corpora are provided below:

\subsection{IEMOCAP}

\vspace{-1mm}

The IEMOCAP database \cite{busso2008iemocap} was collected using multi-modal sensors that capture motion, audio, and video of acted human interactions. The corpus contains 10,039 utterances from ten subjects (five male and five female) who target expressing categorical emotions. In addition, the utterances are divided into improvised and scripted conditions where the speakers use utterances from a fixed script in the latter case. In this work, follow the suggestion from \cite{zhang2018attention} and focus on the improvised sessions. 

\vspace{-1mm}

\subsection{CREMA-D}

The CREMA-D \cite{cao2014crema} corpus is a multi-modal database of emotional speech collected from 91 actors, 48 of whom are male, and 43 are female. The set contains 7,442 speech recordings that simulate emotional expressions, including happy, sad, anger, fear, and neutral. 

\vspace{-1mm}

\subsection{MSP-Improv}

The MSP-Improv~\cite{busso2016msp} corpus was created to study naturalistic emotions captured from improvised scenarios. The corpus includes audio and visual data of utterances spoken in natural condition (2,785 utterances), target condition (652 target utterances in an improvised scenario), improvised condition (4,381 utterances from the remainder of the improvised scenarios), and read speech condition (620 utterances). The data is collected from 12 participants (six male and six female). Similar to the IEMOCAP data set, we use the data only from the improvised conditions.

\begin{table}
    
    \centering
    \caption{Statistics of emotion labels in three different SER data sets.}
    
    \begin{tabular}{p{1.25cm}p{1.25cm}p{1.25cm}p{1.25cm}p{1.25cm}p{1.25cm}}

        \toprule
        \multicolumn{1}{c}{} & 
        \multicolumn{1}{c}{\textbf{Neutral}} & 
        \multicolumn{1}{c}{\textbf{Happy}} &
        \multicolumn{1}{c}{\textbf{Sad}} &
        \multicolumn{1}{c}{\textbf{Angry}} & 
        \multicolumn{1}{c}{\textbf{All}}
        \rule{0pt}{2ex} \\ \midrule

        \multicolumn{1}{l}{\textbf{IEMOCAP}} & 
        \multicolumn{1}{c}{1099} &
        \multicolumn{1}{c}{947} &
        \multicolumn{1}{c}{608} &
        \multicolumn{1}{c}{289} &
        \multicolumn{1}{c}{2943} \rule{0pt}{2ex} \\
        
        \multicolumn{1}{l}{\textbf{CREAM-D}} & 
        \multicolumn{1}{c}{1972} &
        \multicolumn{1}{c}{1219} &
        \multicolumn{1}{c}{588} &
        \multicolumn{1}{c}{1019} &
        \multicolumn{1}{c}{4798} \rule{0pt}{2ex} \\
        
        \multicolumn{1}{l}{\textbf{MSP-Improv}} & 
        \multicolumn{1}{c}{2072} &
        \multicolumn{1}{c}{1184} &
        \multicolumn{1}{c}{739} &
        \multicolumn{1}{c}{585} &
        \multicolumn{1}{c}{4580} \rule{0pt}{2ex} \\
        
        \bottomrule
    \end{tabular}

    \label{tab:dataset}
\end{table}

\section{Problem setup}

In this section, we describe preliminaries and the problem setup of the attack framework. To improve readability, we summarize the notations adopted in this paper in  \autoref{tab:notation}.

\begin{table}
    \centering
    \caption{Notation used in this paper.}
    \begin{tabular}{p{0.5cm}p{0.5cm}}
        \toprule

        \multicolumn{1}{l}{$\mathbf{D_{p}}$} & 
        \multicolumn{1}{l}{Training data set of the private model.} \\
        
        \multicolumn{1}{l}{$\mathbf{D_{s_{1}}}, ..., \mathbf{D_{s_{m}}}$} & 
        \multicolumn{1}{l}{Training data set of the shadow model.} \\
        
        \multicolumn{1}{l}{$\mathbf{D_{a}}$} & 
        \multicolumn{1}{l}{Training data set of the attack model.} \\
        
        \midrule
        
        \multicolumn{1}{l}{$\mathbf{M_{p}}$} & 
        \multicolumn{1}{l}{Private training model.} \\
        
        \multicolumn{1}{l}{$\mathbf{M_{s_{1}}}, ..., \mathbf{D_{s_{m}}}$} & 
        \multicolumn{1}{l}{Shadow models, mimic private training model.} \\
        
        \multicolumn{1}{l}{$\mathbf{M_{a}}$} & 
        \multicolumn{1}{l}{Attack model.} \\
        
        \midrule
        \multicolumn{1}{l}{$\mathbf{X}$} & 
        \multicolumn{1}{l}{Speech data.} \\
        
        \multicolumn{1}{l}{$\mathbf{y}$} & 
        \multicolumn{1}{l}{Emotion label.} \\
        
        \multicolumn{1}{l}{$\mathbf{z}$} & 
        \multicolumn{1}{l}{Sensitive attribute label.} \\
        
        \multicolumn{1}{l}{$t$} & 
        \multicolumn{1}{l}{Global training epoch in FL.} \\
        
        \multicolumn{1}{l}{$k$} & 
        \multicolumn{1}{l}{Client index.} \\
        
        \multicolumn{1}{l}{$\mathbf{g}_{k}^{t}$} & 
        \multicolumn{1}{l}{Model gradients of $k$-th client at $t$-th global training epoch.} \\
        
        \multicolumn{1}{l}{$\mathbf{\theta}^{t}$} & 
        \multicolumn{1}{l}{Global SER model parameters at $t$-th global training epoch.} \\
        
        \multicolumn{1}{l}{$\mathbf{\theta}_{k}^{t}$} & 
        \multicolumn{1}{l}{SER model parameter updates of $k$-th client at $t$-th global} \\
        
        \multicolumn{1}{l}{} & 
        \multicolumn{1}{l}{training epoch.} \\
        
        \multicolumn{1}{l}{$\mathbf{\nabla W_{i}}$} & 
        \multicolumn{1}{l}{$i$-th layer's weight updates of the SER model.} \\
        
        \multicolumn{1}{l}{$\mathbf{\nabla b_{i}}$} & 
        \multicolumn{1}{l}{$i$-th layer's bias updates of the SER model.} \\
        
        \multicolumn{1}{l}{$\mathbf{\psi}$} & 
        \multicolumn{1}{l}{Attack model parameters.} \\
        
        \multicolumn{1}{l}{$\eta$} & 
        \multicolumn{1}{l}{Learning rate.} \\

        \bottomrule
    \end{tabular}
    \label{tab:notation}
\end{table}

\begin{figure*}
	\centering
	\includegraphics[width=\linewidth]{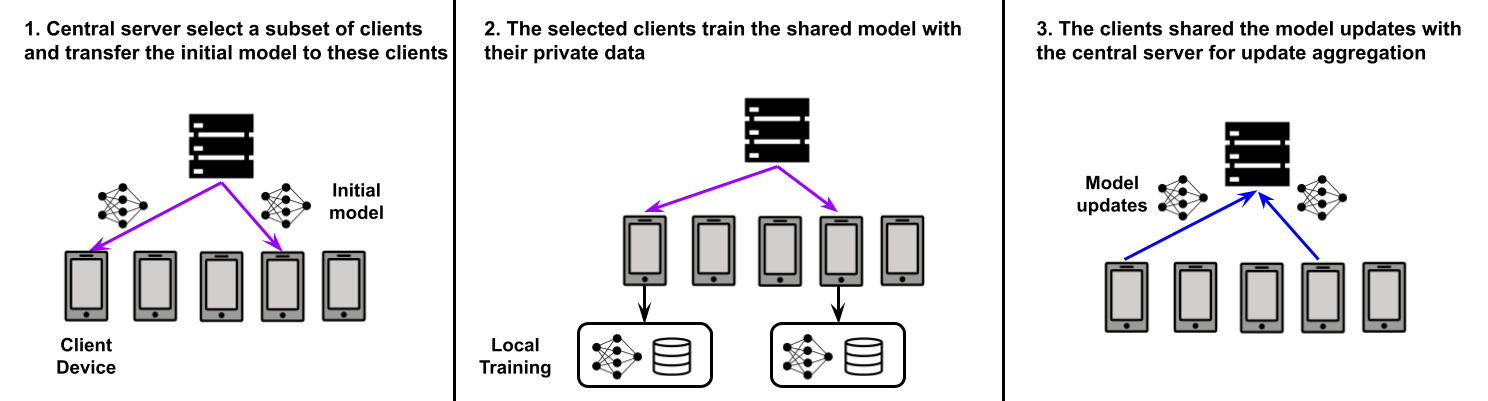}
    \caption{The figure shows the training process of a global round in Federated Learning. The server shares a global model with several clients to perform local training with their private data in each global round. Eventually, the clients transfer the model updates (model parameters/gradients) to the central server for aggregating the latest global model. (Image credit: OpenMoji \cite{openmoji})}
    \label{fig:fl_framework}
\end{figure*}

\subsection{Federated Learning}

Federated learning is a training algorithm that enables multiple clients to collaboratively train a joint ML model, coordinated through a central server. For example, in a typical FL training round shown in Fig~\ref{fig:fl_framework}, a subset of selected clients receive a global model, which they locally train with their private data. Afterward, the clients share their model updates (model parameters/gradients) to the central server. Finally, the server aggregates the model updates to obtain the global model for the next training round. FedSGD and FedAvg are two common approaches to produce the aggregated models in FL~\cite{kairouz2019advances}. 

\begin{figure}
	\centering
	\includegraphics[width=\linewidth]{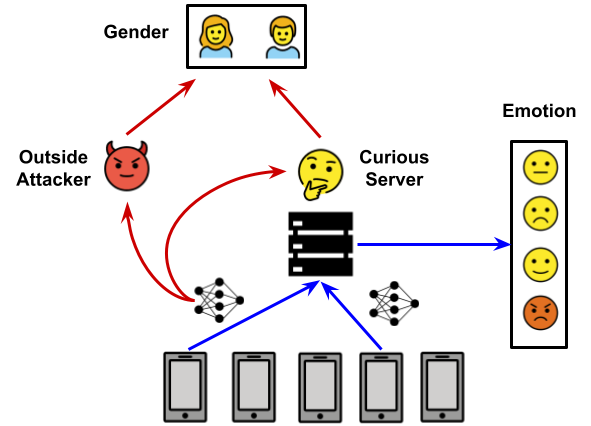}
    \caption{The figure shows the problem setup of the attribute inference attack in this article. We use SER as the primary application in our experiments and in the attack, the adversary (which can be an external attacker or the server itself) attempts to predict the gender using the shared model updates. (Image credit: OpenMoji \cite{openmoji})}
    \label{fig:attacl_problem}
\end{figure}

\vspace{1.5mm}

\noindent \textbf{FedSGD} We define ${\theta}^{t}$ as the global model parameter in $t$-th global round. In FedSGD, the $k$-th client locally computes gradient updates $\mathbf{{g}_{k}^{t}}$ based on one batch of private training data, and sends $\mathbf{{g}_{k}^{t}}$ to the server. Assuming $K$ clients containing a total of N samples participate in the $t$-th round of training where each client is of sample size $n_{i}$ and learning rate is $\eta$, the server computes the updated global model as:

\begin{equation}
    \mathbf{\theta^{t+1}} = \mathbf{\theta^{t}} - \sum_{k=1}^{K}\frac{n_{k}}{N}{\mathbf{{g}_{k}^{t}}}
\end{equation}

\noindent \textbf{FedAvg} In the FedAvg algorithm, each client locally takes several epochs of model updates using its entire training data set $\mathbf{D}_{k}$ and obtains a local model with parameters $\mathbf{{\theta}_{k}^{t}}$. Each client then submits the resulting model to the server, which calculates a weighted average shown below:

\begin{equation}
    \mathbf{\theta^{t+1}} = \sum_{k}^{K}\frac{n_{k}}{N}{\mathbf{\theta}_{k}^{t}}
\end{equation}

\begin{figure*}
	\centering
	\includegraphics[width=\linewidth]{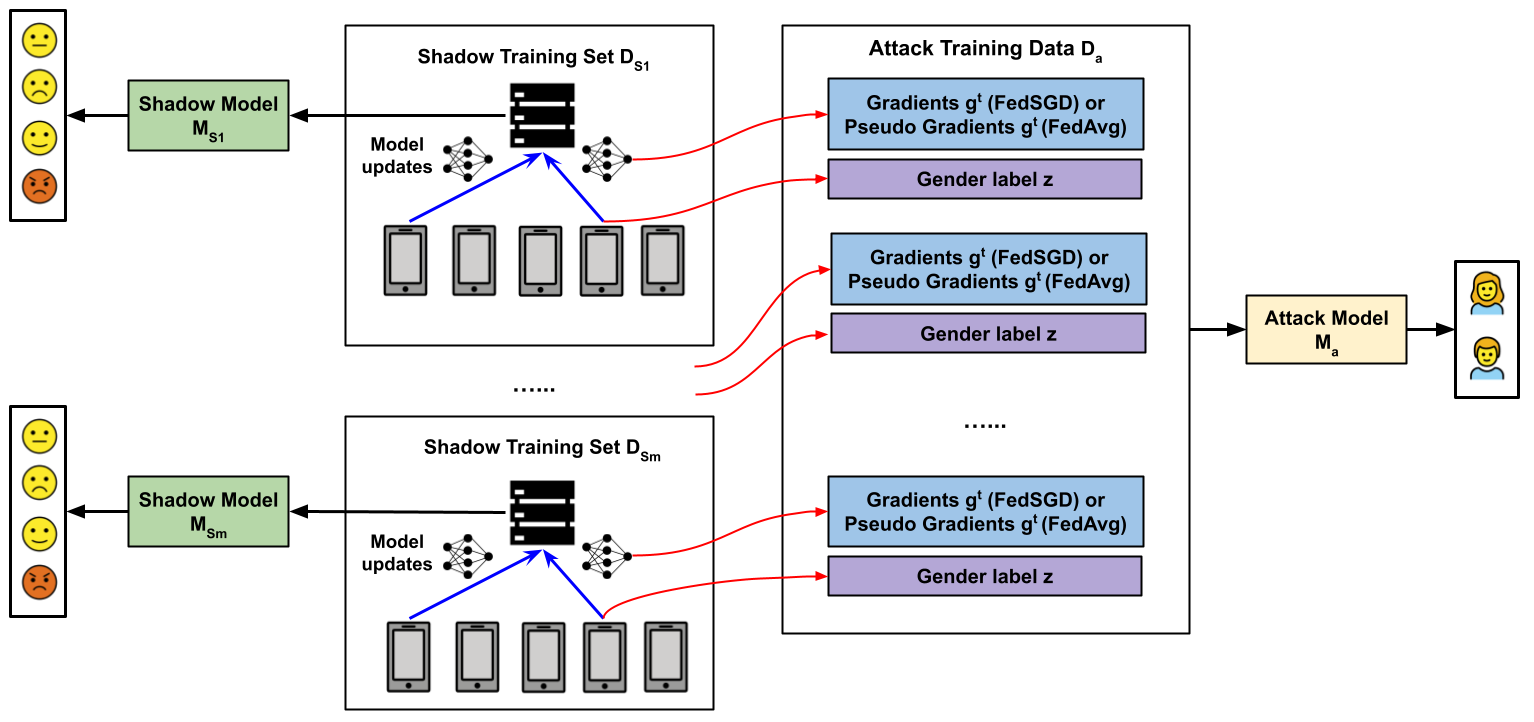}
    \caption{The framework of our proposed attack. We form $m$ shadow training data sets and train $m$ shadow models that mimics the private FL training process. All the model updates during the FL shadow training process are used to construct the attack training data set $D_{a}$. We use $D_{a}$ to train attack model $M_{a}$ in predicting sensitive attributes from clients. (Image credit: OpenMoji \cite{openmoji})}
    \label{fig:attack_framework}
\end{figure*}

\subsection{Problem Definition}

\autoref{fig:attacl_problem} shows the attack problem setup we investigate in this work. In this study, the \textit{primary task} is SER, models for which are trained using the FL framework, while in the \textit{Adversarial task} the attacker attempts to predict the client's gender label. %(\textit{Adversarial task}). 
We follow a setup in which we have a private-labeled data set $\mathbf{D_{p}}$ from a number of clients, where each client has a feature set $\mathbf X$ and an emotion label set $\mathbf y$. We also assume a gender label $z$ associated with each client. This work focuses on the white-box attack, where the attacker knows the model architecture and hyper-parameters like batch size, local epochs, and learning rate. We also assume that the attacker does not have access to the private training data. However, the adversary can access public data-sets with a similar data format to $\mathbf{D_{p}}$. Similar to the attacking framework proposed in \cite{geng2021towards}, we define two attack scenarios based on two FL algorithms: \textbf{FedSGD} and \textbf{FedAvg}.

\vspace{1.25mm}

\noindent \textbf{FedSGD} In the FedSGD framework, we assume that the attacker has access to shared gradients $\mathbf{{g}_{k}^{t}}$ from the $k^\text{th}$ client in the $t^\text{th}$ global training epoch but not the private speech data $\mathbf X_{k}$. The attacker attempts to predict the sensitive attribute $z_{k}$ (e.g. gender label) of the $k$-th client using $\mathbf{{g}_{k}^{t}}$.

\vspace{1.25mm}

\noindent \textbf{FedAvg} In the FedAvg framework, the attacker has access to the global model parameter $\mathbf{\theta^{t}}$ and shared model parameters $\mathbf{\theta_{k}^{t}}$ from $k$-th client at the $t$-th global training round but not the private speech data $\mathbf X_{k}$. The attacker's goal is to infer the sensitive attribute $z_{k}$ (e.g. gender label) of the $k$-th client using $\mathbf{\theta^{t}}$ and $\mathbf{\theta_{k}^{t}}$.

\section{Attacking Formulation}

In this section, we describe our proposed method for attribute inference attack in detail. Our attack focuses on training a classification model using the model updates generated in the FL setting; either the model gradients $\mathbf{{g}}$ or the model parameters $\mathbf{\theta}$, to infer the sensitive label $\mathbf{z}$ associated with a client. Our proposed attack framework is shown in \autoref{fig:attack_framework}. In each subsequent subsection, we explain each step in more detail. Here is the summary of the steps when IEMOCAP data set is used as $\mathbf{D_{p}}$:

\vspace{1mm}
\begin{enumerate}[leftmargin=0.5cm]
    
    \item \textbf{Service provider}: Private training of the SER models using the FL setup using IEMOCAP data set as $\mathbf{D_{p}}$.
    \vspace{1mm}
    \item \textbf{Attacker}: 
    
    \begin{enumerate}[leftmargin=0.5cm]
        \item The attacker trains SER models using datasets available to them (denoted as $\mathbf{D_s}$) such that the training mimics the private training setup in a client. In our experiments we set $\mathbf{D_s}$ to be CREMA-D and MSP-Improv datasets. 
        
        \item Collect the shared model updates during the shadow training in step a) to generate an attack data set $\mathbf{D_{a}}$.
        
        \item The attacker trains a gender classification model $\mathbf{M_{a}}$ using $\mathbf{D_{a}}$. 
        
        \item Finally, the attacker infers the gender of the clients in $\mathbf{D_{p}}$ using the shared model updates from the private training and $\mathbf{M_{a}}$.
    \end{enumerate}

\end{enumerate}

\subsection{Private Training}

We refer our target SER model training to obtain $\mathbf{M_{p}}$ as private training. In this paper, the private training is done in the FL setting, where we have a private training data set $D_{p}$ with emotion labels $\mathbf{y_{p}}$. All the private training data are on the client's device and  not accessible by the central server. The server performs the FL using two algorithms: FedSGD and FedAvg. As described above, only the model gradients or the model parameters are shared with the central aggregator in the FedSGD algorithm and the FedAvg algorithm, respectively. In this work, we also assume that the attacker cannot access the private training data. However, the shared training updates (either the gradients or the model parameters) are insecure, where the attacker can obtain this information.

\subsection{Shadow Training}

The shadow training is first proposed in the membership inference attack~\cite{shokri2017membership}. In this paper, we use a similar attack framework to construct our attribute inference attack. Specifically, the shadow models $\mathbf{M_{s_{1}}}, \mathbf{M_{s_{2}}}, ..., \mathbf{M_{s_{m}}}$ are trained to mimic the private training model $\mathbf{M_{p}}$. In this paper, the shadow models also aim to classify emotion categories from speech features. To train the shadow models, the attacker typically collects a set of shadow training data sets. The objective of the attacker is to collect shadow datasets as similar in format and distribution to the private dataset, as possible. While the individual shadow data sets may overlap, the private training data set and the shadow data set may not overlap. In each experiment, we propose to use one SER data set (e.g., IEMOCAP) as the private training data set and two other SER data sets (e.g., CREMA-D and MSP-Improv) as the shadow training data set. The shadow models are to predict emotion categories similar to the private model. Since we are focusing on the white-box attack, we train the shadow models in a similar fashion to the private FL training, where the shadow models have the same model architecture as the private model and are with the same hyper-parameters (e.g., learning rate, local epochs) used in the private training. We use 80\% of data to train each shadow SER model. 

\begin{figure*}
	\centering
	\includegraphics[width=\linewidth]{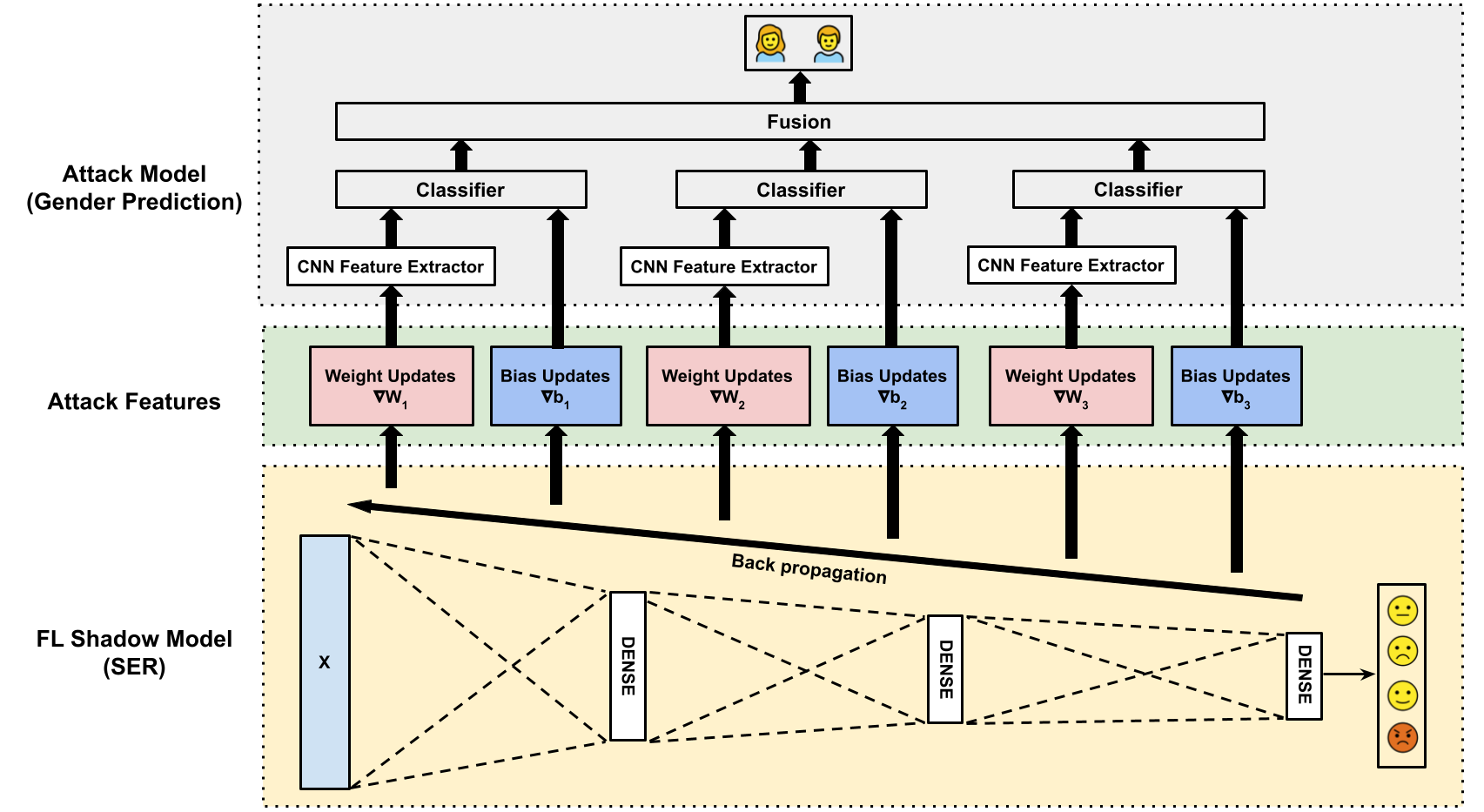}
    \caption{The architecture of our proposed attack model. The model updates (raw gradients in FedSGD; pseudo gradients in FedAvg) are the input features to train the attack model. The CNN feature extractor uses the weight update matrix to compute the hidden representation, which combines with bias updates. Finally, we use the combined attack feature representation to infer the gender attribute. (Image credit: OpenMoji \cite{openmoji})}
    \label{fig:attack_model}
\end{figure*}

\subsection{Attack Model}

In this work, we compose the attack training data set $\mathbf{D_{a}}$ using the shared model updates which are generated during the shadow training experiments. Given a shared model update $\mathbf{g_{k}^{t}}$ from $k^\text{th}$ client in the shadow training process, we record the gender of the k-th client $z_{k}$ as the label of $\mathbf{g_{k}^{t}}$. We then use $\mathbf{D_{a}}$ to train the attack model $\mathbf{M_{a}}$ to infer the gender attribute $\mathbf{z}$. Finally, we define the attack training data under two FL learning scenarios:

\vspace{1.5mm}

\noindent \textbf{FedSGD} In the FedSGD framework, the attack model takes the gradients $\mathbf{g}_{k}^{t}$ as the input data to train the attack model in predicting the gender label $z_{k}$. The attacker model attempts to minimize the following cross-entropy loss between model predictions $\mathbf{M_a}(\mathbf{g_k^t}, \mathbf{\psi})$ and the labels $z_{k}$. $\mathbf{\psi}$ are the parameters of the model $\mathbf{M_a}$. with parameters $\mathbf{\psi}$: 

\begin{equation}
    \min_{\psi} \;\; L(M_{a}(\mathbf{g}_{k}^{t}; \mathbf{\psi}), z_{k})
\end{equation}

\noindent \textbf{FedAvg} Herein, only the global model parameters $\mathbf{\theta^{t}}$ and the updated model parameters $\mathbf{\theta^{t}_{k}}$ from the k-th client are accessible by the attacker but not the raw gradients. Thus, we derive a pseudo gradient that is similar to previous work in \cite{geng2021towards} as the attack model's input data. Specifically, we assume that the global model undergoes $T$ times of local updates at the $k$-th client, where $T$ is the product between the local training epoch and the number of mini-batches within a local training epoch. Thus, we can define the following with the \textbf{pseudo gradients} $\mathbf{g'}_{k}^{t}$ and the learning rate $\eta$:

\begin{equation}
    \mathbf{\theta^{t}_{k}} = \mathbf{\theta^{t}} - T \eta \cdot \mathbf{g'}_{k}^{t}
\end{equation}

Eventually, we can define the \textbf{pseudo gradients} $\mathbf{g'}_{k}^{t}$ as:

\begin{equation}
    \mathbf{g'}_{k}^{t}  = \frac{1}{T\eta} (\mathbf{\theta^{t}} - \mathbf{\theta^{t}_{k}})
\end{equation}

Given this, we aim to train the attack model with parameters $\psi$ to minimize the following cross entropy loss function:

\begin{equation}
    \min_{\psi} \;\; L(M_{a}(\mathbf{g'}_{k}^{t}; \mathbf{\psi}), z_{k})
\end{equation}

Our attack model is similar to the membership inference attack model architecture in \cite{nasr2019comprehensive}. The attack model consists of CNN feature extractors and classifiers as shown in \autoref{fig:attack_model}. $\mathbf{\nabla W_{i}}$ and $\mathbf{\nabla b_{i}}$ represent the weight updates and the bias updates in $\mathbf{g}$ corresponding to the $i^\text{th}$ layer, respectively. Each layer's weight updates (generated from the FL training) is first fed into a three-layer CNN feature extractor to compute the hidden representation. We then flatten the output from the CNN feature extractor and concatenate it with the layer's bias updates. We then pass this combined representation to the MLP classifier to predict gender. We use a fusion layer to combine the predictions from the individual layer classifiers; the fusion method used in this work is a weighted average function. We determine the importance of each layer's gender prediction output based on the size of the shared updates. Finally, we evaluate the performance of the attack model using the shared model updates generated in the private FL setting where the attack model's goal is to infer the gender labels of clients in the private training data set.

\begin{table*}

    \centering
    \caption{Prediction results of the SER model trained under the FedSGD and FedAvg FL scenarios. The accuracy and unweighted average recall (UAR) scores of the SER task on individual data set as the private training data set are reported. The feature sets include Emo-Base obtained from the OpenSMILE toolkit and APC, Vq-APC, Tera, NPC, and DeCoAR 2.0 computed using pre-trained models.}

    \begin{tabular}{ccccccccccccc}
        
        \toprule
        & 
        \multicolumn{6}{c}{\textbf{FedSGD}} &
        \multicolumn{6}{c}{\textbf{FedAvg}}
        \rule{0pt}{2.25ex} \\ \cmidrule(lr){2-7} \cmidrule(lr){8-13}

        \textbf{Speech Feature} & 
        \multicolumn{2}{c}{\textbf{IEMOCAP}$\mathbf{(D_{p})}$} &
        \multicolumn{2}{c}{\textbf{CREMA-D}$\mathbf{(D_{p})}$} &
        \multicolumn{2}{c}{\textbf{MSP-Improv}$\mathbf{(D_{p})}$} &
        \multicolumn{2}{c}{\textbf{IEMOCAP}$\mathbf{(D_{p})}$} &
        \multicolumn{2}{c}{\textbf{CREMA-D}$\mathbf{(D_{p})}$} &
        \multicolumn{2}{c}{\textbf{MSP-Improv}$\mathbf{(D_{p})}$}
        \rule{0pt}{2.25ex} \\ % \cline{2-13}

        &
        {\textbf{Acc}} & 
        {\textbf{UAR}} &
        {\textbf{Acc}} & 
        {\textbf{UAR}} &
        {\textbf{Acc}} & 
        {\textbf{UAR}} &
        {\textbf{Acc}} & 
        {\textbf{UAR}} &
        {\textbf{Acc}} & 
        {\textbf{UAR}} &
        {\textbf{Acc}} & 
        {\textbf{UAR}}
        \rule{0pt}{2.25ex} \\ \cmidrule(lr){1-1} \cmidrule(lr){2-3} \cmidrule(lr){4-5} \cmidrule(lr){6-7} \cmidrule(lr){8-9} \cmidrule(lr){10-11} \cmidrule(lr){12-13}
        
        \textbf{Emo-Base} &
        62.94\% & 
        61.62\% & 
        69.81\% & 
        64.24\% & 
        54.17\% & 
        47.23\% & 
        61.22\% & 
        61.25\% & 
        71.26\% & 
        66.38\% & 
        54.37\% & 
        47.67\% 
        \rule{0pt}{2.25ex} \\
        
        \textbf{APC} &
        64.71\% & 
        63.27\% & 
        71.78\% & 
        69.08\% & 
        56.81\% & 
        51.17\% & 
        64.22\% & 
        63.34\% & 
        72.66\% & 
        68.48\% & 
        55.85\% & 
        49.41\% 
        \rule{0pt}{2.25ex} \\
        
        \textbf{Vq-APC} &
        62.05\% & 
        62.07\% & 
        71.73\% & 
        68.62\% & 
        56.24\% & 
        50.93\% & 
        63.59\% & 
        62.50\% & 
        72.80\% & 
        69.32\% & 
        56.54\% & 
        50.67\% 
        \rule{0pt}{2.25ex} \\
        
        \textbf{NPC} &
        64.34\% & 
        63.18\% & 
        72.11\% & 
        67.84\% & 
        54.79\% & 
        47.87\% & 
        63.63\% & 
        61.75\% & 
        73.25\% & 
        68.24\% & 
        56.14\% & 
        48.99\% 
        \rule{0pt}{2.25ex} \\
        
        \textbf{DeCoAR 2.0} &
        \textbf{65.99\%} & 
        \textbf{65.18\%} & 
        \textbf{75.88\%} & 
        \textbf{73.69\%} & 
        57.42\% & 
        52.48\% & 
        \textbf{65.67\%} & 
        \textbf{64.21\%} & 
        \textbf{76.77\%} & 
        \textbf{71.64\%} & 
        \textbf{57.99\%} & 
        \textbf{53.87\%} 
        \rule{0pt}{2.25ex} \\
        
        \textbf{Tera} &
        62.19\% & 
        62.05\% & 
        72.91\% & 
        70.31\% & 
        \textbf{57.77\%} & 
        \textbf{52.60\%} & 
        62.50\% & 
        61.31\% & 
        73.50\% & 
        68.16\% & 
        57.54\% & 
        52.67\% 
        \rule{0pt}{2.25ex} \\
        
        \bottomrule

    \end{tabular}
    
    \label{tab:fl_result}
\end{table*}

\section{Experiments}

In this section, we describe our experimental setup including data processing, data setup, and training details. The implementation of this paper is at \href{https://github.com/usc-sail/fed-ser-leakage}{https://github.com/usc-sail/fed-ser-leakage}.

\subsection{Data Preprocessing}

To investigate the effectiveness of the proposed attack framework, we train our SER models on a variety of speech representations. We first generate the Emo-Base feature set using the OpenSMILE toolkit \cite{eyben2010opensmile} for each utterance. In addition to the knowledge-based speech feature set, we propose to evaluate our framework on SUPERB (Speech Processing Universal PERformance Benchmark) \cite{yang21c_interspeech}, which is designed to provide a standard and comprehensive testbed for pre-trained models on various downstream speech tasks. We compute the deep speech representations from the pre-trained models that are available in SUPERB including APC \cite{chung2019unsupervised}, Vq-APC \cite{chung2020vqapc}, Tera \cite{liu2021tera}, NPC \cite{liu2020nonautoregressive}, and DeCoAR 2.0 \cite{ling2020decoar}. We further compute the global average of the last layer's hidden state as the final feature from the pre-trained model's output. Using the last hidden state is suggested in prior works for downstream tasks \cite{chung2019unsupervised, liu2020nonautoregressive, ling2020decoar, Liu_2020}. Our feature sizes are 988 in Emo-Base; 512 in APC, Vq-APC, and NPC; 768 in Tera and DeCoAR 2.0.

We apply z-normalization to the speech features within each speaker. Since there are only 10 speakers in the IEMOCAP data set and 12 speakers in the MSP-Improv data set, we further divide each speaker's data in these two data sets into $10$ parts of equal size. This mimics a scenario where a single person owns multiple clients and their data is distributed across them (e.g. a person can own cell phones, tables and computer devices across wich their data is distributed). This division is to create more clients for the FL training. Each divided speaker data is the local training data on a client. In the CREMA-D data set, each speaker is a unique client in the FL training as there are 91 unique speakers in the dataset. We leave 20\% of speakers as the test data. Then, we repeat the experiments 5 times with test folds of different speakers. Finally, we report the average results of the 5-fold experiments.

\subsection{Data setup}

%In this paper, 
We simulate the experiments using different private training data sets. For instance, in the case of the IEMOCAP data set being the private training data set $\mathbf{D_{p}}$, the MSP-Improv data set and CREMA-D data set are  combined to train shadow models $\mathbf{M_{s_{1}}, ..., M_{s_{m}}}$. In our experiments, we set $m=5$. Next, we compose the attack training data set using the shared model updates generated from the FL shadow training process and use it to train the attack model $\mathbf{M_{a}}$. In the above example, we evaluate the performance of $\mathbf{M_{a}}$ using the model updates generated in the FL that uses IEMOCAP data set as the private training data $\mathbf{D_{p}}$. Finally, we repeat the same experiments with the MSP-Improv data set and the CREMA-D data set as the private training data set $\mathbf{D_{p}}$. We also run the experiment on each speech representation.

\vspace{-2mm}

\subsection{Model and training Details}

In this work, we use the multilayer perceptron (MLP) as the SER model architecture. The model consists of 2 dense layers with layer sizes in \{256, 128\}. We choose ReLU as the activation function and the dropout rate as 0.2. We implement both FedSGD and FedAvg algorithms in training the SER model. Only 10\% of the clients participate in each global round. 80\% of the data at a client is for local training, and the rest 20\% is for validation. We set the local training batch size as 20. Specifically, we set the learning rate as 0.0005 and the local training epoch as 1 in the FedAvg algorithm. On the other hand, we set the learning rate as 0.1 in the FedSGD algorithm for faster convergence. The total global training epochs are 200 in both FL scenarios. In the shadow training, the attacker designs the shadow model with the same architecture and trains the shadow models with the same hyper-parameters used in the private FL training. The attacker collects the shared model updates and then uses these data to train the attack model in predicting gender label $z$. We set the learning rate as 0.0001 in training the attack model. The attack model consists of CNN feature extractors with 3 CNN layers. Each subsequent CNN layer consists of \{16, 32, 64\} filters of size $5 \times 5$. Maxpooling is also applied after each CNN layer. We choose ReLU as the activation function and the dropout rate of \{0.1, 0.2\} in the attack model.

\subsection{Evaluation Metrics}

\noindent \textbf{Speech Emotion Recognition} We use the Unweighted Average Recall (UAR) score to evaluate predictions in SER models. 

\noindent \textbf{Attacker Task} Inspired by \cite{feng2022user}, \cite{wainakh2021user}, and \cite{hu2021source}, we define the \textit{attack success rate (ASR)} to measure the attacker performance. Specifically, given a client in FL training setup, we randomly select a gradient update from the whole training process, and make a prediction. We repeat this process 10 times for each client, and report the percentage of correct predictions as the attack success rate. We average the attack success rate from all clients as the final performance of the attacker. More formally, the \textit{ASR} over $K$ clients is defined as:

\begin{equation}
    \textit{ASR} = \sum_{i}^{K}\frac{\text{(\# of successful attacks)}_{i}}{\text{(\# of total attacks)}_{i}}
\end{equation}

\begin{table}

    \centering
    \caption{Prediction results of the attribute inference attack model. The attack success rate (ASR) of the gender prediction task is reported. The attack frameworks are simulated in two FL scenarios. Bold indicates the best attack performance in one data combination.}

    \begin{tabular}{ccccccc}
        
        \toprule
        
        $\mathbf{D_{p}}$ & 
        $\mathbf{D_{s}}$ & 
        \multicolumn{1}{c}{\textbf{Feature}} &
        \multicolumn{1}{c}{\textbf{FedSGD}} &
        \multicolumn{1}{c}{\textbf{FedAvg}}
        \rule{0pt}{2.25ex} \\ \cmidrule(lr){1-1} \cmidrule(lr){2-2} \cmidrule(lr){3-3} \cmidrule(lr){4-4} \cmidrule(lr){5-5}
        
        \multirow{6}{*}{\textbf{IEMOCAP}} & 
        & 
        \textbf{Emo-Base} & 
        89.33\% & 
        87.45\% & 
        \rule{0pt}{2.25ex} \\
        
        & 
        & 
        \textbf{APC} & 
        89.95\% & 
        86.00\% & 
        \rule{0pt}{2.25ex} \\
        
        & 
        {\textbf{MSP-Improv}} & 
        \textbf{Vq-APC} & 
        81.98\% & 
        78.70\% & 
        \rule{0pt}{2.25ex} \\
        
        & 
        {\textbf{CREMA-D}} & 
        \textbf{NPC} & 
        94.67\% & 
        93.73\% & 
        \rule{0pt}{2.25ex} \\
        
        & 
        & 
        \textbf{DeCoAR 2.0} & 
        87.05\% & 
        86.87\% & 
        \rule{0pt}{2.25ex} \\
        
        & 
        & 
        \textbf{Tera} & 
        87.07\% & 
        90.50\% & 
        \rule{0pt}{2.25ex} \\
        
        \midrule
        \multirow{6}{*}{\textbf{CREMA-D}} & 
        & 
        \textbf{Emo-Base} & 
        73.87\% & 
        81.98\% & 
        \rule{0pt}{2.25ex} \\
        
        & 
        & 
        \textbf{APC} & 
        83.63\% & 
        83.90\% & 
        \rule{0pt}{2.25ex} \\
        
        & 
        {\textbf{IEMOCAP}} & 
        \textbf{Vq-APC} & 
        73.38\% & 
        70.72\% & 
        \rule{0pt}{2.25ex} \\
        
        & 
        {\textbf{MSP-Improv}} & 
        \textbf{NPC} & 
        82.19\% & 
        81.39\% & 
        \rule{0pt}{2.25ex} \\
        
        & 
        & 
        \textbf{DeCoAR 2.0} & 
        76.97\% & 
        70.42\% & 
        \rule{0pt}{2.25ex} \\
        
        & 
        & 
        \textbf{Tera} & 
        86.67\% & 
        82.12\% & 
        \rule{0pt}{2.25ex} \\
        
        \midrule
        \multirow{6}{*}{\textbf{MSP-Improv}} & 
        & 
        \textbf{Emo-Base} & 
        91.36\% & 
        82.86\% & 
        \rule{0pt}{2.25ex} \\
        
        & 
        & 
        \textbf{APC} & 
        94.21\% & 
        88.04\% & 
        \rule{0pt}{2.25ex} \\
        
        & 
        {\textbf{IEMOCAP}} & 
        \textbf{Vq-APC} & 
        88.45\% & 
        80.25\% & 
        \rule{0pt}{2.25ex} \\
        
        & 
        {\textbf{CREMA-D}} & 
        \textbf{NPC} & 
        96.90\% & 
        93.72\% & 
        \rule{0pt}{2.25ex} \\
        
        & 
        & 
        \textbf{DeCoAR 2.0} & 
        95.02\% & 
        90.53\% & 
        \rule{0pt}{2.25ex} \\
        
        & 
        & 
        \textbf{Tera} & 
        93.30\% & 
        86.87\% & 
        \rule{0pt}{2.25ex} \\
        
        \bottomrule

    \end{tabular}
    
    \label{tab:attack_result}
\end{table}

\section{Results}

In this section, we present SER results on different private data sets. We also show results of the attack model in predicting gender labels of the clients in the private data set.

\begin{table*}

    \centering
    \caption{Prediction results of the attribute inference attack model using shared updates between different layers in the SER model. The attack success rate (ASR) of the gender prediction task are reported. The attack frameworks are simulated in two FL scenarios.}
    \begin{tabular}{ccccccccc}
        
        \toprule
        \multirow{2}{*}{$\mathbf{D_{p}}$} &
        \multirow{2}{*}{$\mathbf{D_{s}}$} &
        \multirow{2}{*}{\textbf{Feature}} & 
        
        \multicolumn{3}{c}{\textbf{FedSGD (ASR)}} &
        \multicolumn{3}{c}{\textbf{FedAvg (ASR)}}
        \rule{0pt}{2.25ex} \\ \cmidrule(lr){4-6} \cmidrule(lr){7-9}
        
        & 
        & 
        &
        \multicolumn{1}{c}{$\mathbf{\nabla W_{1}}$ + $\mathbf{\nabla b_{1}}$} &
        \multicolumn{1}{c}{$\mathbf{\nabla W_{2}}$ + $\mathbf{\nabla b_{2}}$} &
        \multicolumn{1}{c}{$\mathbf{\nabla W_{3}}$ + $\mathbf{\nabla b_{3}}$} &
        \multicolumn{1}{c}{$\mathbf{\nabla W_{1}}$ + $\mathbf{\nabla b_{1}}$} &
        \multicolumn{1}{c}{$\mathbf{\nabla W_{2}}$ + $\mathbf{\nabla b_{2}}$} &
        \multicolumn{1}{c}{$\mathbf{\nabla W_{3}}$ + $\mathbf{\nabla b_{3}}$}
        \rule{0pt}{2.25ex} \\ \cmidrule(lr){1-1} \cmidrule(lr){2-2} \cmidrule(lr){3-3} \cmidrule(lr){4-4} \cmidrule(lr){5-5} \cmidrule(lr){6-6} \cmidrule(lr){7-7} \cmidrule(lr){8-8} \cmidrule(lr){9-9}
        
        \multirow{6}{*}{\textbf{IEMOCAP}} & 
        & 
        \textbf{Emo-Base} & 
        89.35\% & 
        60.22\% & 
        57.90\% & 
        87.38\% & 
        59.20\% & 
        55.27\% 
        \rule{0pt}{2.25ex} \\
        
        & 
        & 
        \textbf{APC} & 
        89.88\% & 
        69.25\% & 
        58.50\% & 
        86.25\% & 
        55.90\% & 
        52.10\% 
        \rule{0pt}{2.25ex} \\
        
        & 
        {\textbf{MSP-Improv}} & 
        \textbf{Vq-APC} & 
        81.82\% & 
        62.15\% & 
        55.35\% & 
        78.70\% & 
        58.88\% & 
        54.02\% 
        \rule{0pt}{2.25ex} \\
        
        & 
        {\textbf{CREMA-D}} & 
        \textbf{NPC} & 
        94.75\% & 
        76.10\% & 
        58.23\% & 
        93.70\% & 
        59.52\% & 
        51.30\% 
        \rule{0pt}{2.25ex} \\
        
        & 
        & 
        \textbf{DeCoAR 2.0} & 
        86.95\% & 
        68.20\% & 
        58.45\% & 
        86.78\% & 
        56.67\% & 
        54.30\% 
        \rule{0pt}{2.25ex} \\
        
        & 
        & 
        \textbf{Tera} & 
        87.05\% & 
        73.78\% & 
        59.25\% & 
        90.38\% & 
        58.65\% & 
        54.48\% 
        \rule{0pt}{2.25ex} \\
        
        \midrule
        \multirow{6}{*}{\textbf{CREMA-D}} & 
        & 
        \textbf{Emo-Base} & 
        73.82\% & 
        59.91\% & 
        58.02\% & 
        81.78\% & 
        59.28\% & 
        56.12\% 
        \rule{0pt}{2.25ex} \\
        
        & 
        & 
        \textbf{APC} & 
        83.68\% & 
        66.19\% & 
        55.61\% & 
        83.54\% & 
        59.87\% & 
        57.97\% 
        \rule{0pt}{2.25ex} \\
        
        & 
        {\textbf{IEMOCAP}} & 
        \textbf{Vq-APC} & 
        73.19\% & 
        62.80\% & 
        55.21\% & 
        70.17\% & 
        61.07\% & 
        58.84\% 
        \rule{0pt}{2.25ex} \\
        
        & 
        {\textbf{MSP-Improv}} & 
        \textbf{NPC} & 
        81.62\% & 
        71.27\% & 
        61.54\% & 
        81.53\% & 
        55.71\% & 
        58.73\% 
        \rule{0pt}{2.25ex} \\
        
        & 
        & 
        \textbf{DeCoAR 2.0} & 
        77.25\% & 
        64.76\% & 
        62.78\% & 
        70.50\% & 
        59.22\% & 
        53.63\% 
        \rule{0pt}{2.25ex} \\
        
        & 
        & 
        \textbf{Tera} & 
        86.56\% & 
        69.37\% & 
        62.48\% & 
        82.26\% & 
        55.70\% & 
        55.19\% 
        \rule{0pt}{2.25ex} \\
        
        \midrule
        \multirow{6}{*}{\textbf{MSP-Improv}} & 
        & 
        \textbf{Emo-Base} & 
        91.34\% & 
        60.22\% & 
        56.24\% & 
        83.08\% & 
        52.33\% & 
        52.13\% 
        \rule{0pt}{2.25ex} \\
        
        & 
        & 
        \textbf{APC} & 
        94.15\% & 
        74.81\% & 
        61.26\% & 
        88.42\% & 
        54.80\% & 
        51.06\% 
        \rule{0pt}{2.25ex} \\
        
        & 
        {\textbf{IEMOCAP}} & 
        \textbf{Vq-APC} & 
        88.10\% & 
        68.94\% & 
        63.22\% & 
        80.40\% & 
        56.07\% & 
        53.59\% 
        \rule{0pt}{2.25ex} \\
        
        & 
        {\textbf{CREMA-D}} & 
        \textbf{NPC} & 
        96.84\% & 
        80.60\% & 
        65.06\% & 
        94.03\% & 
        52.90\% & 
        51.68\% 
        \rule{0pt}{2.25ex} \\
        
        & 
        & 
        \textbf{DeCoAR 2.0} & 
        94.95\% & 
        74.78\% & 
        63.32\% & 
        90.61\% & 
        57.24\% & 
        52.14\% 
        \rule{0pt}{2.25ex} \\
        
        & 
        & 
        \textbf{Tera} & 
        93.26\% & 
        75.53\% & 
        61.48\% & 
        87.05\% & 
        54.38\% & 
        49.52\% 
        \rule{0pt}{2.25ex} \\
        
        \bottomrule
        
    \end{tabular}
    \vspace{-1.5mm}
    \label{tab:attack_diff_layer}
\end{table*}

\subsection{Speech Emotion Recognition using FL}

The emotion prediction results of FL training on different private training data sets are shown in \autoref{tab:fl_result}. We report the SER prediction results in accuracy (Acc) scores and unweighted average recall (UAR) scores. We observe that the knowledge-based feature set, Emo-Base, performs comparably in the SER task to prior works \cite{satt2017efficient, ramet2018context} that use spectrograms or MFCCs. In addition, we can observe that the deep speech representation, DeCoAR 2.0, yields the best UAR scores in prediction emotions when the private training data sets are CREMA-D (FedSGD: 73.69\%; FedAvg: 71.64\%) and IEMOCAP (FedSGD: 65.18\%; FedAvg: 64.21\%). TERA feature set produces the best UAR scores when the private training data is MSP-Improv (FedSGD: 52.60\%). Our results show that the SER task performs better when the training data sets are IEMOCAP and CREMA-D. In summary, these results suggest that our SER models, trained within a FL architecture, produce reasonable predictions for the SER task.

\subsection{Attribute Inference Attack}

\autoref{tab:attack_result} summarizes the performance of the proposed attack model. We report the attack success rate of the attack model (gender inference). Given a speech representation, we simulate the attack frameworks in three data combinations with different SER data set as the private training data set ($\mathbf{D_{p}}$). From the table, we observe that the attack model can consistently use just the client's shared model updates, to infer the client's gender information.

\vspace{1mm}

\noindent \textbf{Speech features:} We observe that this attribute inference attack is possible regardless of the speech representation (UAR scores are all above $70\%$) used for the SER task. It is also interesting to note that the attack model yields the best overall performance in predicting gender labels when the deep speech representations, such as APC and Tera, are the input data to the SER task but not the knowledge-based feature set, Emo-Base. Noticeably, these deep speech representations also provide the best overall emotion prediction performance as shown in \autoref{tab:fl_result}. Typically, deep speech representations are more generalized feature embeddings for downstream speech applications. Besides the knowledge-based speech feature set, Tera and APC, we find that other deep speech representations can also generate shared model updates in the FL, which can leak significant attribute information about the client. 

\vspace{1mm}

\noindent \textbf{FedSGD and FedAvg:} We find that this attribute information leakage exists in both these FL learning algorithms. Increasingly, we discover that the attack model has higher chances to predict the client's gender information when we train the private SER model using the FedSGD algorithm. One reason behind this is the model updates in FedAvg create averaged model differences for the inference attack,   This observation is consistent with the results of data reconstruction attacks reported in \cite{geng2021towards}.

\vspace{1mm}

\noindent \textbf{Data Set:} In general, we find that this attribute inference attack is possible with all data set combinations used in this work. However, the attack model appears to have slightly better gender prediction performance when the private training data sets are either the IEMOCAP or the MSP-Improv. This is probably because the CREMA-D data set consists of more unique speakers, creating more diverse FL training updates. 

\vspace{1mm}

\noindent \textbf{Summary:} The experimental results above demonstrate that our proposed attack framework is robust to infer gender information of the clients involved in the FL without accessing the client's private speech feature data but the shared model updates (raw gradients or pseudo gradients). Consequently, even though the speech feature samples are not accessible by the attacker in FL when training the global SER model, the attribute information about a client can leak through the model updates.

\begin{table*}

    \centering
    \caption{Prediction results of the attribute attacker model and SER model using different dropout probability. The unweighted average recall (UAR) scores and the attack success rates (ASR) for SER task and attacker task are reported, respectively. The attack frameworks are simualted in two FL scenarios.}
    \begin{tabular}{ccccccccc}
        
        \toprule
        
        \multirow{2}{*}{$\mathbf{D_{p}}$} &
        \multirow{2}{*}{$\mathbf{D_{s}}$} &
        \multirow{2}{*}{\textbf{Feature}} & 
        
        \multicolumn{3}{c}{\textbf{SER Model (UAR)}} &
        \multicolumn{3}{c}{\textbf{Attack Model (ASR)}}
        \rule{0pt}{2.25ex} \\ \cmidrule(lr){4-6} \cmidrule(lr){7-9}
        
        & 
        & 
        &
        {\textbf{Dropout=0.2}} & 
        {\textbf{Dropout=0.4}} &
        {\textbf{Dropout=0.6}} & 
        {\textbf{Dropout=0.2}} &
        {\textbf{Dropout=0.4}} & 
        {\textbf{Dropout=0.6}}
        \rule{0pt}{2.25ex} \\ \cmidrule(lr){1-1} \cmidrule(lr){2-2} \cmidrule(lr){3-3} \cmidrule(lr){4-4} \cmidrule(lr){5-5} \cmidrule(lr){6-6} \cmidrule(lr){7-7} \cmidrule(lr){8-8} \cmidrule(lr){9-9}
        
        \multirow{6}{*}{\textbf{IEMOCAP}} & 
        & 
        \textbf{Emo-Base} & 
        89.35\% & 
        60.22\% & 
        57.90\% & 
        87.38\% & 
        59.20\% & 
        55.27\% 
        \rule{0pt}{2.25ex} \\
        
        & 
        & 
        \textbf{APC} & 
        89.88\% & 
        69.25\% & 
        58.50\% & 
        86.25\% & 
        55.90\% & 
        52.10\% 
        \rule{0pt}{2.25ex} \\
        
        & 
        {\textbf{MSP-Improv}} & 
        \textbf{Vq-APC} & 
        81.82\% & 
        62.15\% & 
        55.35\% & 
        78.70\% & 
        58.88\% & 
        54.02\% 
        \rule{0pt}{2.25ex} \\
        
        & 
        {\textbf{CREMA-D}} & 
        \textbf{NPC} & 
        94.75\% & 
        76.10\% & 
        58.23\% & 
        93.70\% & 
        59.52\% & 
        51.30\% 
        \rule{0pt}{2.25ex} \\
        
        & 
        & 
        \textbf{DeCoAR 2.0} & 
        86.95\% & 
        68.20\% & 
        58.45\% & 
        86.78\% & 
        56.67\% & 
        54.30\% 
        \rule{0pt}{2.25ex} \\
        
        & 
        & 
        \textbf{Tera} & 
        87.05\% & 
        73.78\% & 
        59.25\% & 
        90.38\% & 
        58.65\% & 
        54.48\% 
        \rule{0pt}{2.25ex} \\
        
        \midrule
        \multirow{6}{*}{\textbf{CREMA-D}} & 
        & 
        \textbf{Emo-Base} & 
        73.82\% & 
        59.91\% & 
        58.02\% & 
        81.78\% & 
        59.28\% & 
        56.12\% 
        \rule{0pt}{2.25ex} \\
        
        & 
        & 
        \textbf{APC} & 
        83.68\% & 
        66.19\% & 
        55.61\% & 
        83.54\% & 
        59.87\% & 
        57.97\% 
        \rule{0pt}{2.25ex} \\
        
        & 
        {\textbf{IEMOCAP}} & 
        \textbf{Vq-APC} & 
        73.19\% & 
        62.80\% & 
        55.21\% & 
        67.17\% & 
        61.07\% & 
        58.84\% 
        \rule{0pt}{2.25ex} \\
        
        & 
        {\textbf{MSP-Improv}} & 
        \textbf{NPC} & 
        81.62\% & 
        71.27\% & 
        61.54\% & 
        81.53\% & 
        55.71\% & 
        58.73\% 
        \rule{0pt}{2.25ex} \\
        
        & 
        & 
        \textbf{DeCoAR 2.0} & 
        77.25\% & 
        64.76\% & 
        62.78\% & 
        70.50\% & 
        59.22\% & 
        53.63\% 
        \rule{0pt}{2.25ex} \\
        
        & 
        & 
        \textbf{Tera} & 
        86.56\% & 
        69.37\% & 
        62.48\% & 
        82.26\% & 
        55.70\% & 
        55.19\% 
        \rule{0pt}{2.25ex} \\
        
        \midrule
        \multirow{6}{*}{\textbf{MSP-Improv}} & 
        & 
        \textbf{Emo-Base} & 
        91.34\% & 
        60.22\% & 
        56.24\% & 
        83.08\% & 
        52.33\% & 
        52.13\% 
        \rule{0pt}{2.25ex} \\
        
        & 
        & 
        \textbf{APC} & 
        94.15\% & 
        74.81\% & 
        61.26\% & 
        88.42\% & 
        54.80\% & 
        51.06\% 
        \rule{0pt}{2.25ex} \\
        
        & 
        {\textbf{IEMOCAP}} & 
        \textbf{Vq-APC} & 
        88.10\% & 
        68.94\% & 
        63.22\% & 
        80.40\% & 
        56.07\% & 
        53.59\% 
        \rule{0pt}{2.25ex} \\
        
        & 
        {\textbf{CREMA-D}} & 
        \textbf{NPC} & 
        96.84\% & 
        80.60\% & 
        65.06\% & 
        94.03\% & 
        52.90\% & 
        51.68\% 
        \rule{0pt}{2.25ex} \\
        
        & 
        & 
        \textbf{DeCoAR 2.0} & 
        94.95\% & 
        74.78\% & 
        63.32\% & 
        90.61\% & 
        57.24\% & 
        52.14\% 
        \rule{0pt}{2.25ex} \\
        
        & 
        & 
        \textbf{Tera} & 
        93.26\% & 
        75.53\% & 
        61.48\% & 
        87.05\% & 
        54.38\% & 
        49.52\% 
        \rule{0pt}{2.25ex} \\
        
        \bottomrule

    \end{tabular}
    
    \label{tab:attack_dropouts}
\end{table*}

\section{Mitigation Possibilities}
There are protection schemes such as cryptography solutions~\cite{bonawitz2017practical, so2021securing} and the use of trusted execution environments~\cite{hashemi2021byzantine, prakash2020mitigating} for secure aggregation. However, cryptography solutions have a significant performance overhead and they are not scalable to systems with many edge devices. Trusted Execution Environments such as Intel SGX~\cite{costan2016intel} provide private environments for data privacy and computational integrity. However, they are not available on all the data centers. In this section, we present an analysis of potential factors related to the attribute information leakage in the FL of the SER model. We aim to investigate a few mitigation strategies based on these possible information leakage factors. Note that all of these proposed mitigation strategies are software-based solutions with low performance overhead. These methods do not need any special system support. 

\subsection{FedSGD and FedAvg}

As we observe from the \autoref{tab:attack_result}, the attack model performs better when we train the SER model using the FedSGD algorithm. Thus, a straightforward defense is to train the global SER model using the FedAvg algorithm. An additional benefit of using the FedAvg is that it significantly reduces the communication overhead during training. The client can transfer the shared model updates after $T$ times of local training instead of each local mini-batch. The primary reason of why the attack model performs worse in the FedAvg scenario is that the averaged model differences contain less information about the training samples than the raw gradients as shown in \cite{geng2021towards}.

\subsection{The layer position of shared model updates}

As suggested in the previous works \cite{mireshghallah2020shredder, narra2021origami}, most information leakage is related to the early layers in a machine learning model. To evaluate this in our attack scenario, we measure the gender prediction performance of the individual classifier without fusion. \autoref{tab:attack_diff_layer} shows the gender prediction performance by using the shared model updates from different layers in the SER model. From \autoref{tab:attack_diff_layer}, we can observe that the attack model can consistently predict the client's gender label using only the shared model updates between the feature input and first dense layer ($\mathbf{\nabla W_{1}}$ and $\mathbf{\nabla b_{1}}$) of the classifier model. However, the gender prediction performance decreases significantly when using the shared model updates between the first to second dense layer ($\mathbf{\nabla W_{2}}$ and $\mathbf{\nabla b_{2}}$) or between the second to the output layer ($\mathbf{\nabla W_{3}}$ and $\mathbf{\nabla b_{3}}$) of the model. The attack success rate is in the range of $50\%-75\%$ using $\mathbf{\nabla W_{2}}$ and $\mathbf{\nabla b_{2}}$ in most of the experiment setups following the FedSGD, and this performance is around $55\%-65\%$ when using input $\mathbf{\nabla W_{3}}$ and $\mathbf{\nabla b_{3}}$. When training the SER model using the FedAvg, the attacks are much weaker using $\mathbf{\nabla W_{2}}$ + $\mathbf{\nabla b_{2}}$ or $\mathbf{\nabla W_{3}}$ + $\mathbf{\nabla b_{3}}$. Thus, we can conclude that the earlier layer's shared updates leak more information about the client's gender attribute when training the SER model using FL.

\begin{figure*}[ht] {
    \centering
    
    \begin{tikzpicture}
        
        % \node[draw=none, align=center] at (0.34\linewidth, 4.9){\textbf{SER performance}};
        
        \node[draw=none,fill=none] at (0, 3.8){\includegraphics[width=0.475\linewidth]{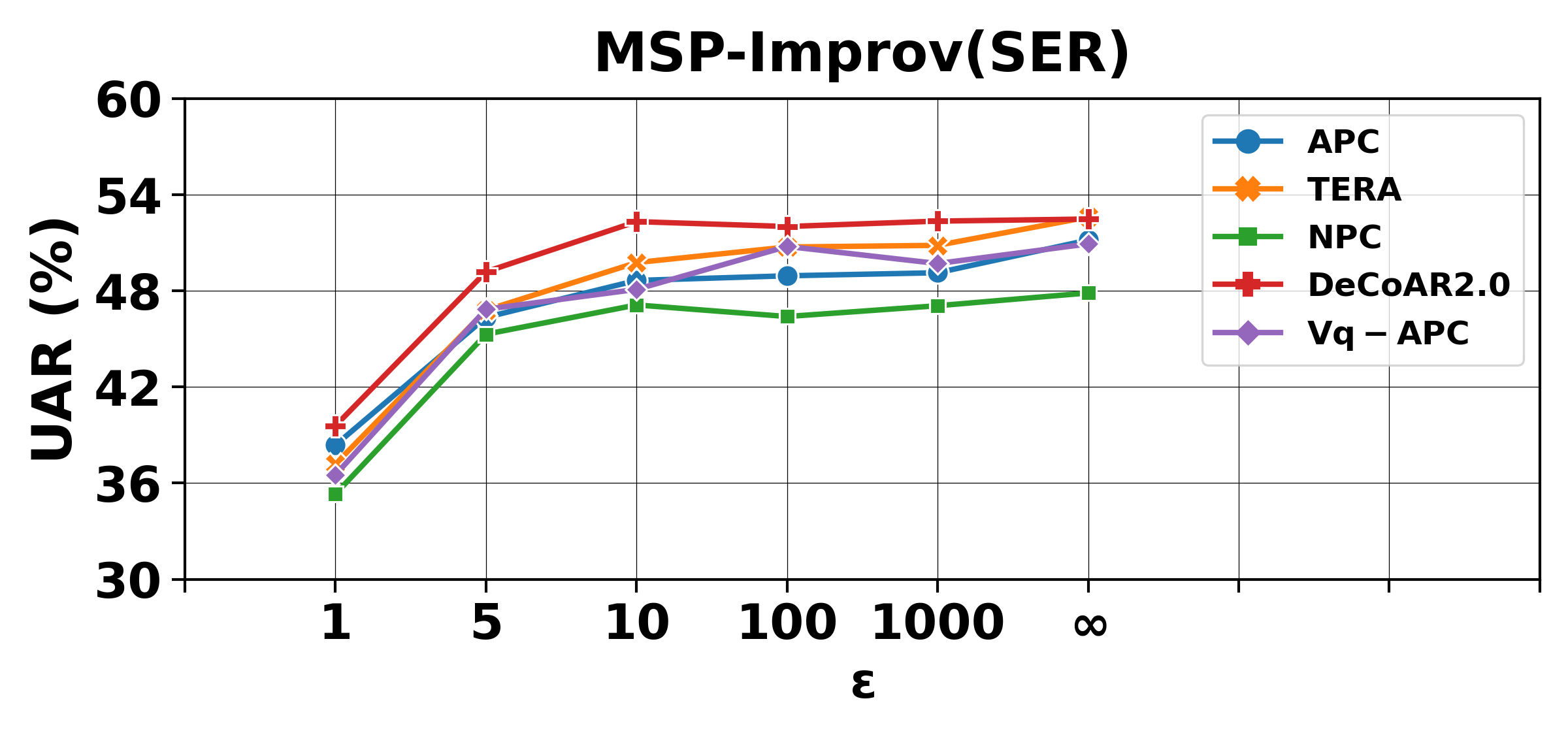}};
        
        \node[draw=none,fill=none] at (0.5\linewidth, 3.8){\includegraphics[width=0.475\linewidth]{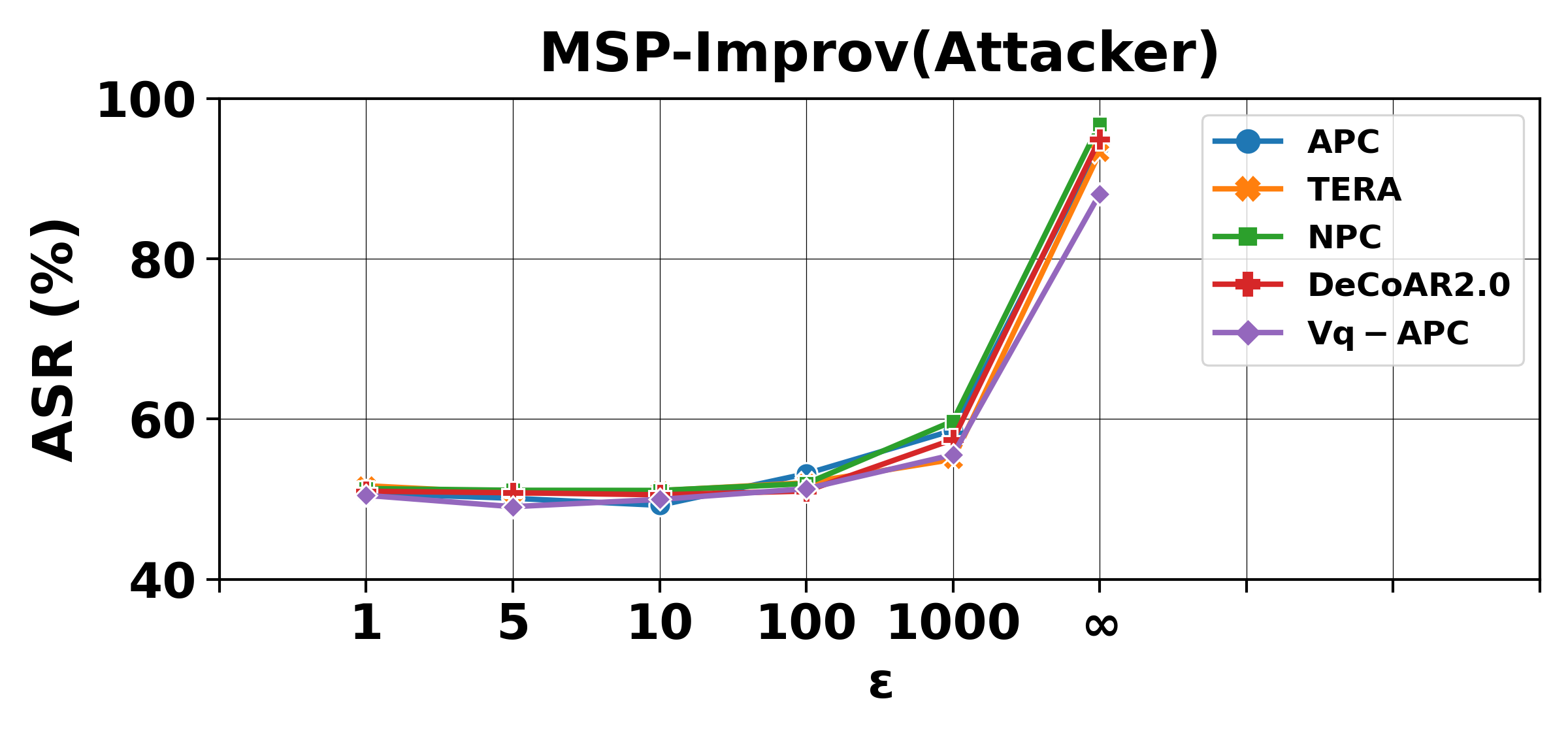}};
        
        \node[draw=none,fill=none] at (0, 7.8){\includegraphics[width=0.475\linewidth]{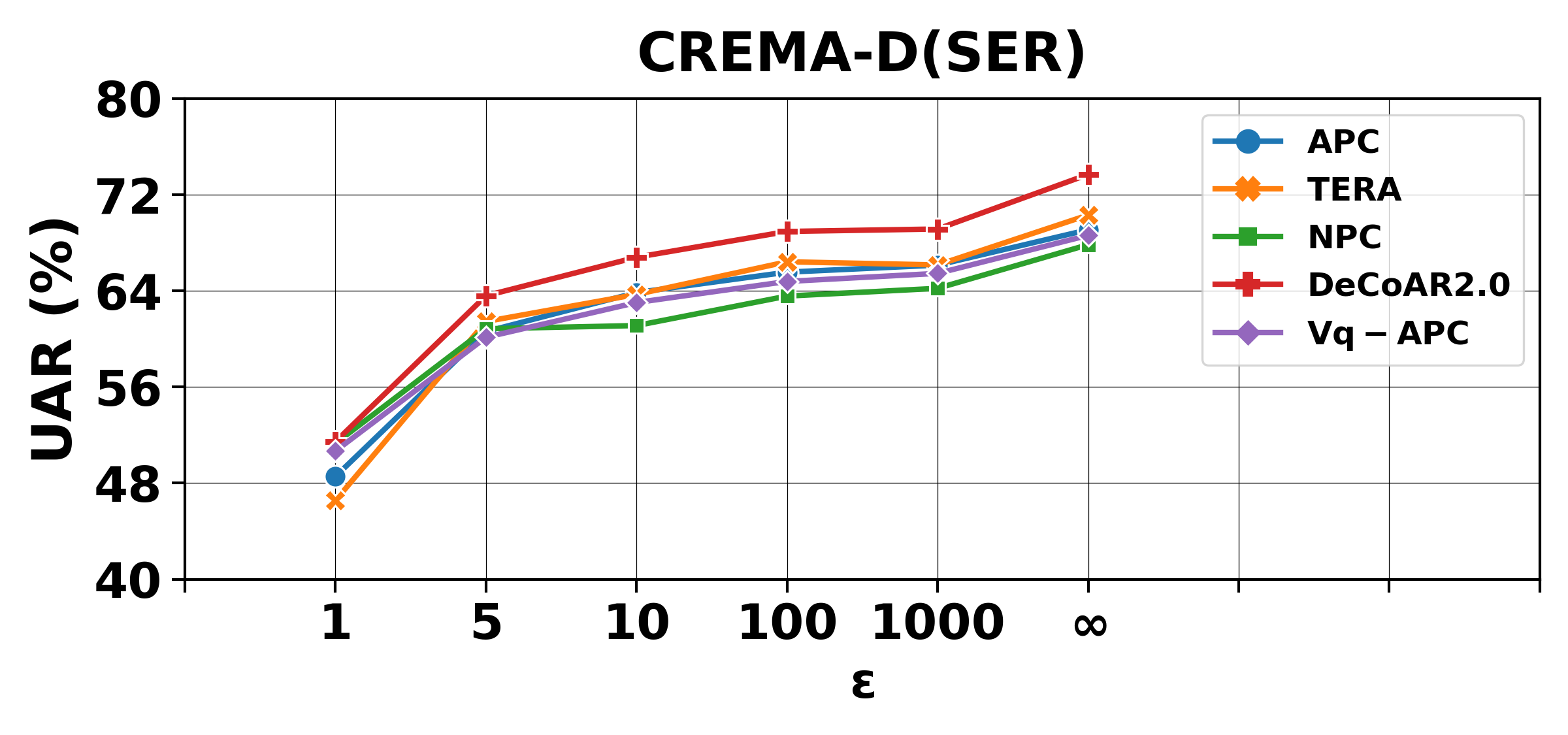}};
        
        \node[draw=none,fill=none] at (0.5\linewidth, 7.8){\includegraphics[width=0.475\linewidth]{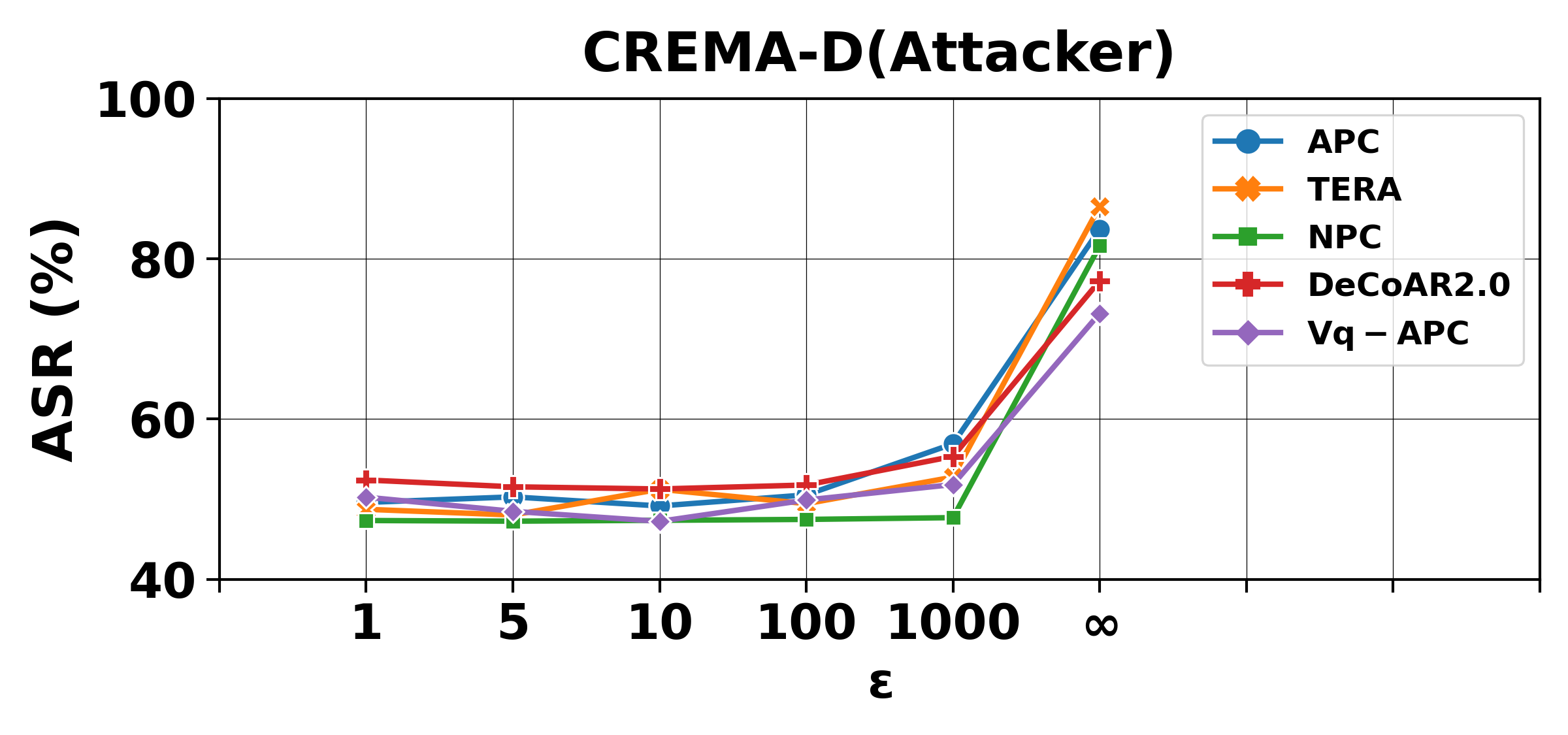}};
        
        \node[draw=none,fill=none] at (0, 11.8){\includegraphics[width=0.475\linewidth]{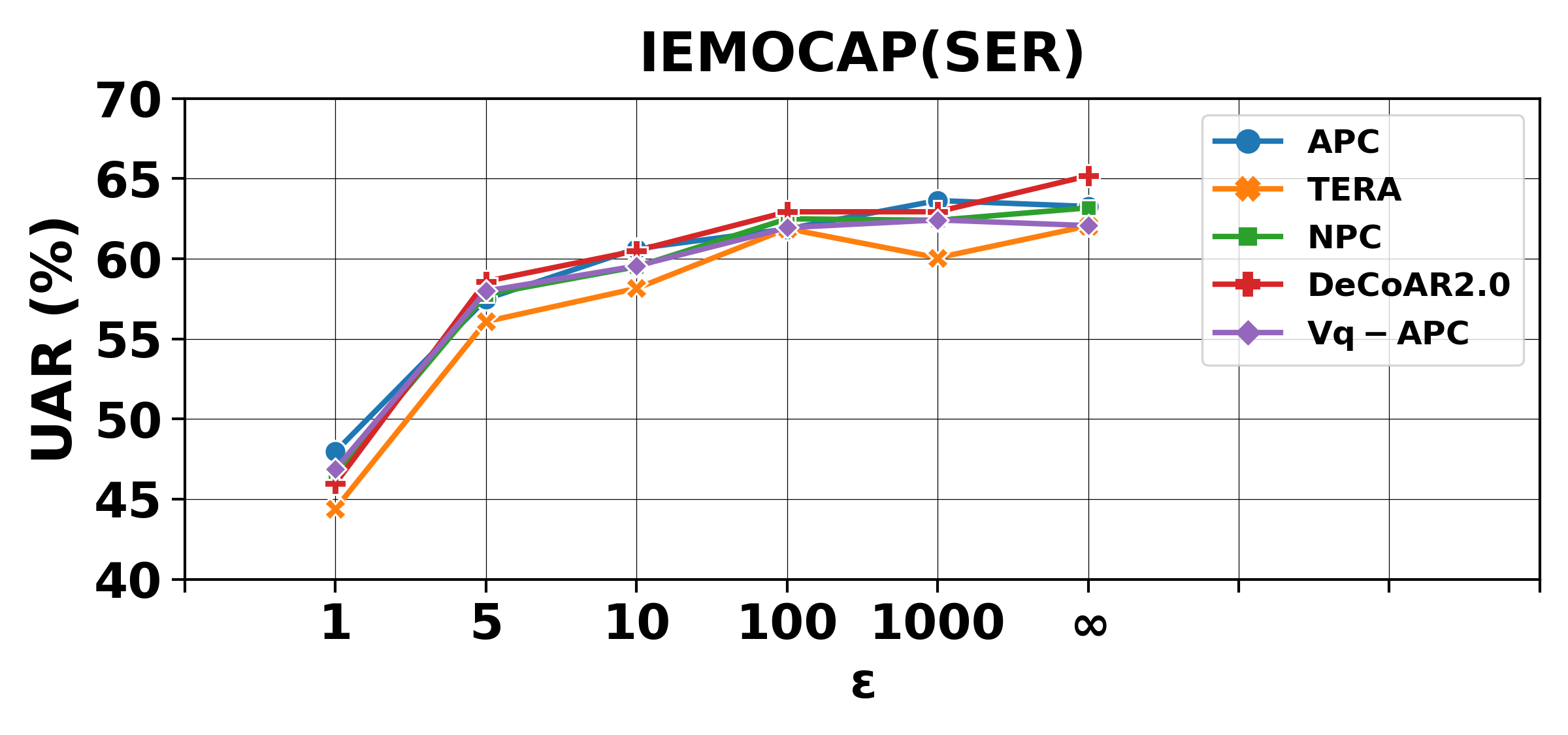}};
        
        \node[draw=none,fill=none] at (0.5\linewidth, 11.8){\includegraphics[width=0.475\linewidth]{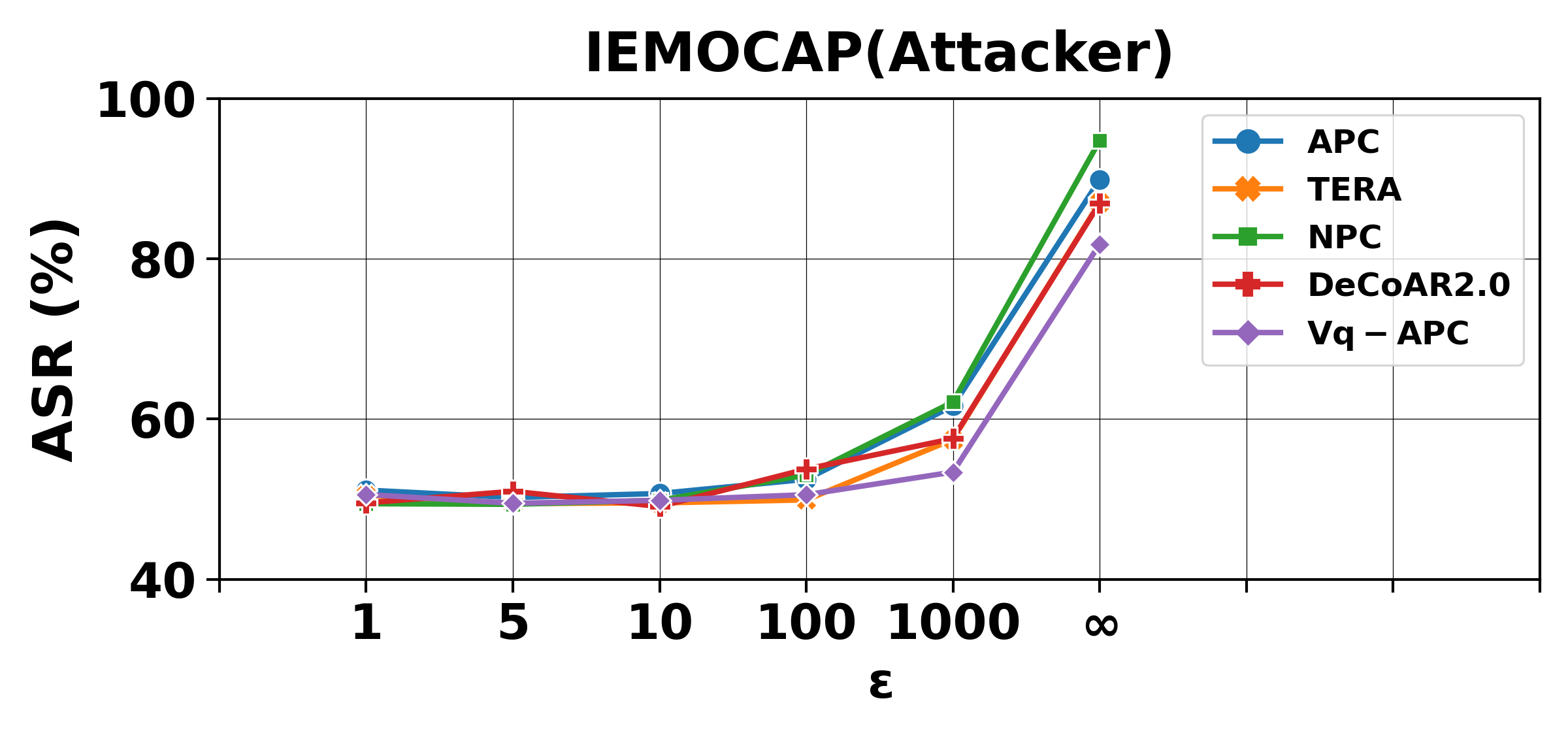}};
        
        % \node[draw=none,fill=none] at (0.66\linewidth, 3.1){\includegraphics[width=0.33\linewidth]{Figures/msp-improv.png}};
        
        % \node[draw=none,fill=none] at (0, 0){\includegraphics[width=0.33\linewidth]{Figures/iemocap.png}};
        
        % \node[draw=none,fill=none] at (0.33\linewidth, 0){\includegraphics[width=0.33\linewidth]{Figures/iemocap.png}};
        
        % \node[draw=none,fill=none] at (0.66\linewidth, 0){\includegraphics[width=0.33\linewidth]{Figures/iemocap.png}};
        
    \end{tikzpicture}
    
    \vspace{-2.5mm}
    \caption{The figure shows the prediction results of the SER task and the attribute inference task at different privacy levels ($\epsilon\in\{1, 5, 10, 100, 1000\}$). A larger $\epsilon$ indicates a weaker privacy guarantee.} 
    \label{fig:attack_dp_results}

    \vspace{-3.75mm}
    
} \end{figure*}

\subsection{Dropout}

Another possible defense is to employ higher dropout \cite{dias2018exploring}, a popular regularization technique used to mitigate overfitting in neural networks. Dropout randomly deactivates activations between neurons, with a probability between 0 and 1. Random de-activations may weaken the attack model because the adversary observes fewer gradients corresponding to the active neurons. We evaluate this assumption by increasing the dropout value to 0.4 and 0.6 after the first dense layer of the MLP classifier. We only increase the dropout rate associated with the first dense layer, since we have shown that this attribute information leakage comes mostly from $\mathbf{\nabla W_{1}}$ and $\mathbf{\nabla b_{1}}$. \autoref{tab:attack_dropouts} shows the UAR scores of the SER task and the inference attack task using the shared model updates, for different dropout values. Increasing dropout value can remove features that is relevant for our primary application, thus decreasing the performance of the SER task. However, our attacks become stronger with increased randomness of dropout applied to the SER model, which is similar to the results shown in \cite{melis2019exploiting}. Our assumption is that there are many shared features which are both informative of emotion and gender. Therefore, removing non-important features for the SER task also eliminates irrelevant features for the gender prediction, while the remaining features are more informative about the gender information.

\subsection{Differential Privacy (DP)}

Differential privacy (DP) prevents information leaks by adding artificial noises. The idea of the DP is to generate data perturbations from different clients to have similar data distribution. More formally, DP perturbs the local data using the mechanism $\mathcal{M}$, such that for neighboring data sets $\mathbf{D}$ and $\mathbf{D'}$, which differ by one sample, we can define the following:

\vspace{-1.5mm}
\begin{definition}[$(\epsilon, \delta)$-DP]
    A random mechanism $\mathcal{M}$ satisfies $(\epsilon, \delta)$-DP, where $\epsilon>0$ and $\delta\in[0, 1)$, if and only if for any two adjacent data sets $\mathbf{D}$ and $\mathbf{D'}$, we have:
    
    \vspace{-1.5mm}
    \begin{equation}
        Pr(\mathcal{M}(\mathbf{D})) \leq e^{\epsilon} Pr(\mathcal{M}(\mathbf{D'})) + \delta
    \end{equation}
\end{definition}
\vspace{-1.5mm}

The parameter $\epsilon>0$ defines the privacy guarantee that the DP provides, and a smaller $\epsilon$ indicates a stronger privacy guarantee. $\delta\in[0, 1)$ indicates the probability that the privacy leaks can occur under the privacy guarantee $\epsilon$~\cite{wei2020federated}. In our recent work \cite{feng2022user}, we have explored using the User-level Differential Privacy (UDP) algorithm to mitigate the attribute inference attack in the FedAvg setup. We extend our prior work to mitigate the attribute inference attack in the FedSGD setting in this work. Specifically, we implement the FedSGD-DP algorithm that has been described in \cite{abadi2016deep}. We experiment with $\epsilon\in\{1, 5, 10, 100, 1000\}$ and $\delta=0.1$. The norm clipping threshold is set to $1.5$. We evaluate the attacker performance using the first layer's model gradients that are similar to our prior work \cite{feng2022user}. \autoref{fig:attack_dp_results} shows the performance of SER model (left column) and the attacker task (right column) under different $\epsilon$. $\epsilon = \infty$ indicates the case where there is no mitigation in the training process. 

\noindent \textbf{SER Performance:} From the SER predictions, we find that SER performance drops by 1-2\% on IEMOCAP and MSP-Improv dataset when $\epsilon=1000$ or $\epsilon=100$. This decrease in SER performance is around 4-5\% on CREMA-D dataset. We also observe that SER performance starts to drop substantially when $\epsilon\leq10$. 

\noindent \textbf{Attacker Performance:} On the other hand, we observe that attacker performance decreases significantly even when $\epsilon$ is at 1000. Similar to SER prediction results, the attacker is unable to perform the privacy attacks when $\epsilon\leq10$.

\begin{figure*}[ht] {
    \centering
    
    \begin{tikzpicture}
        
        \node[draw=none,fill=none] at (0, 2.8){\includegraphics[width=0.25\linewidth]{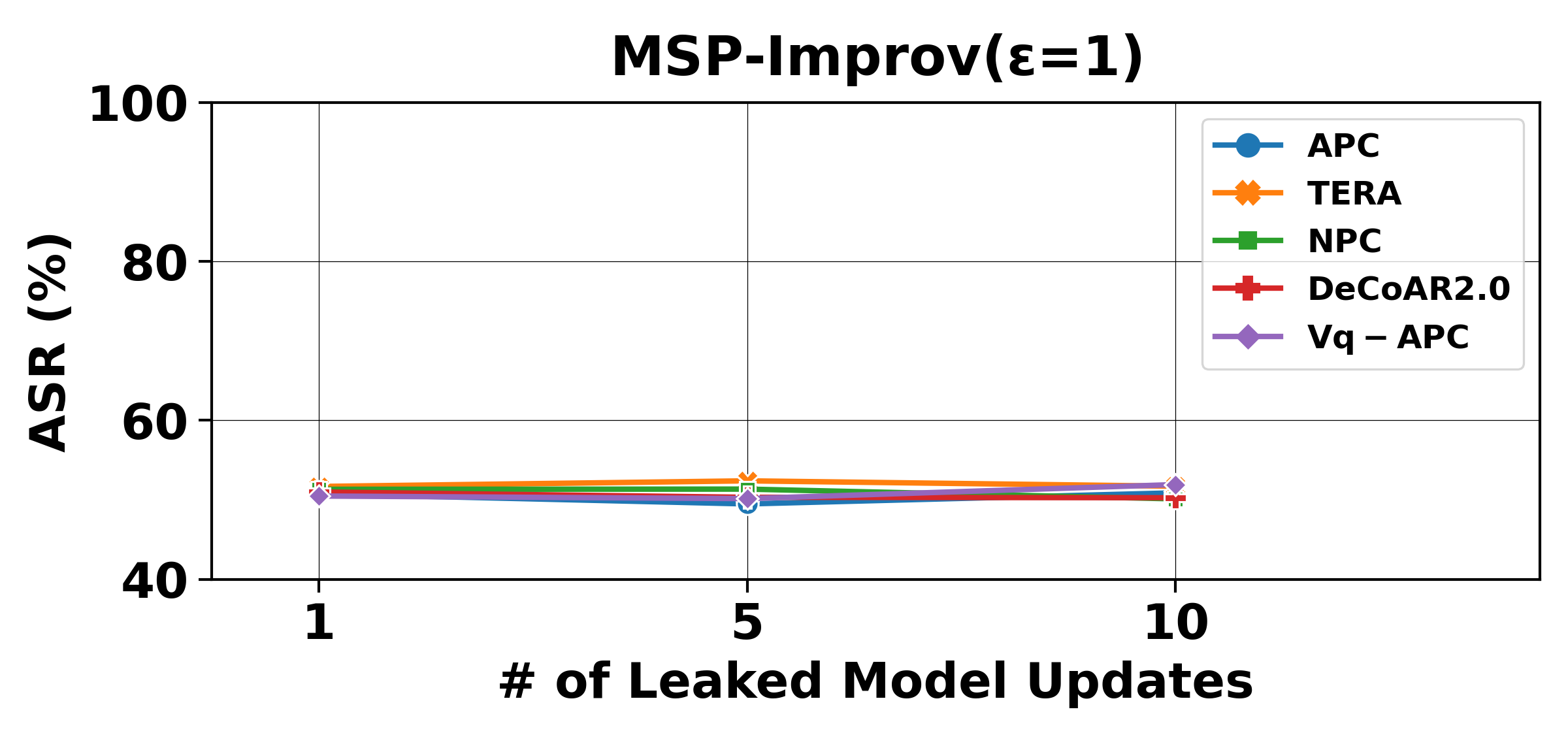}};
        
        \node[draw=none,fill=none] at (0.25\linewidth, 2.8){\includegraphics[width=0.25\linewidth]{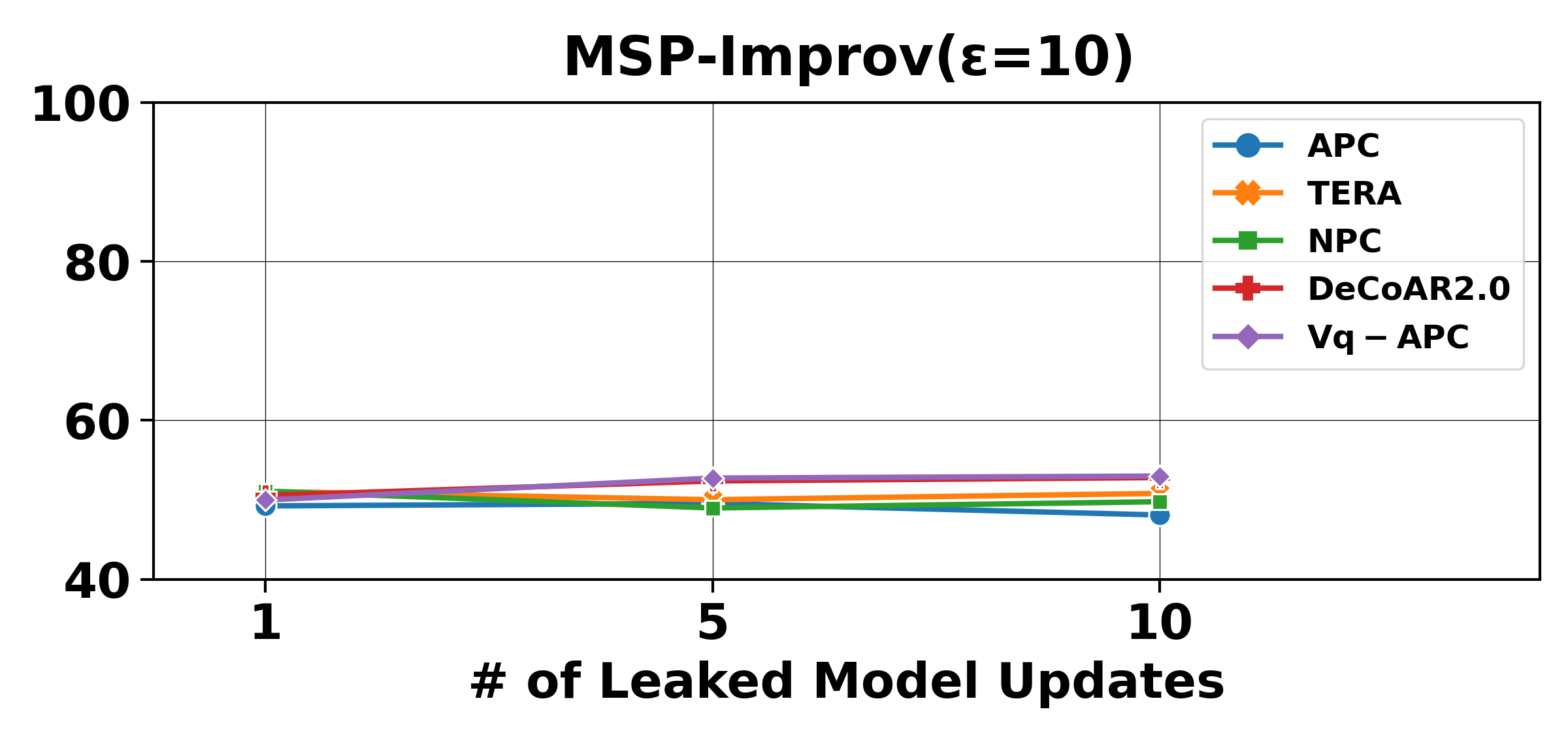}};

        \node[draw=none,fill=none] at (0.5\linewidth, 2.8){\includegraphics[width=0.25\linewidth]{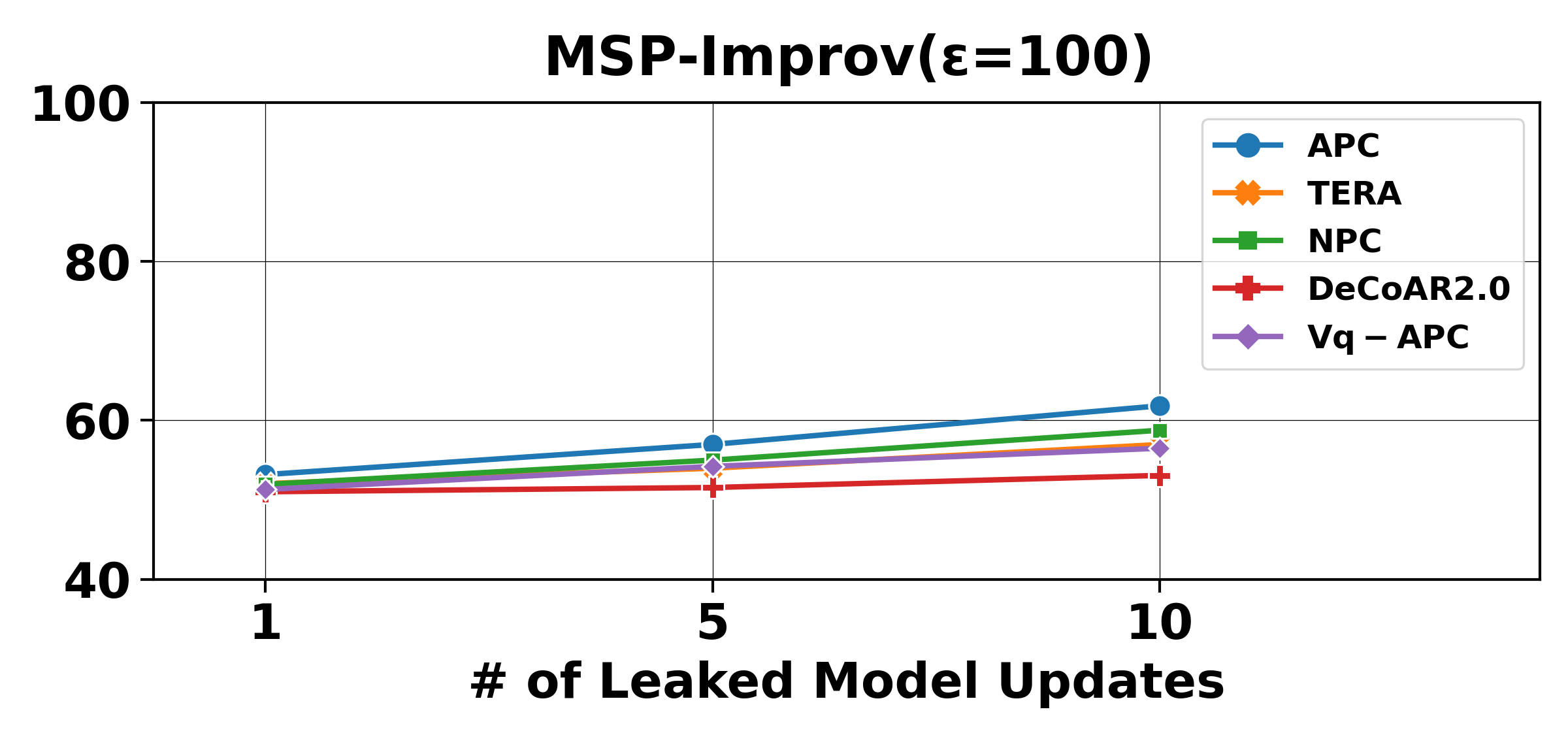}};

        \node[draw=none,fill=none] at (0.75\linewidth, 2.8){\includegraphics[width=0.25\linewidth]{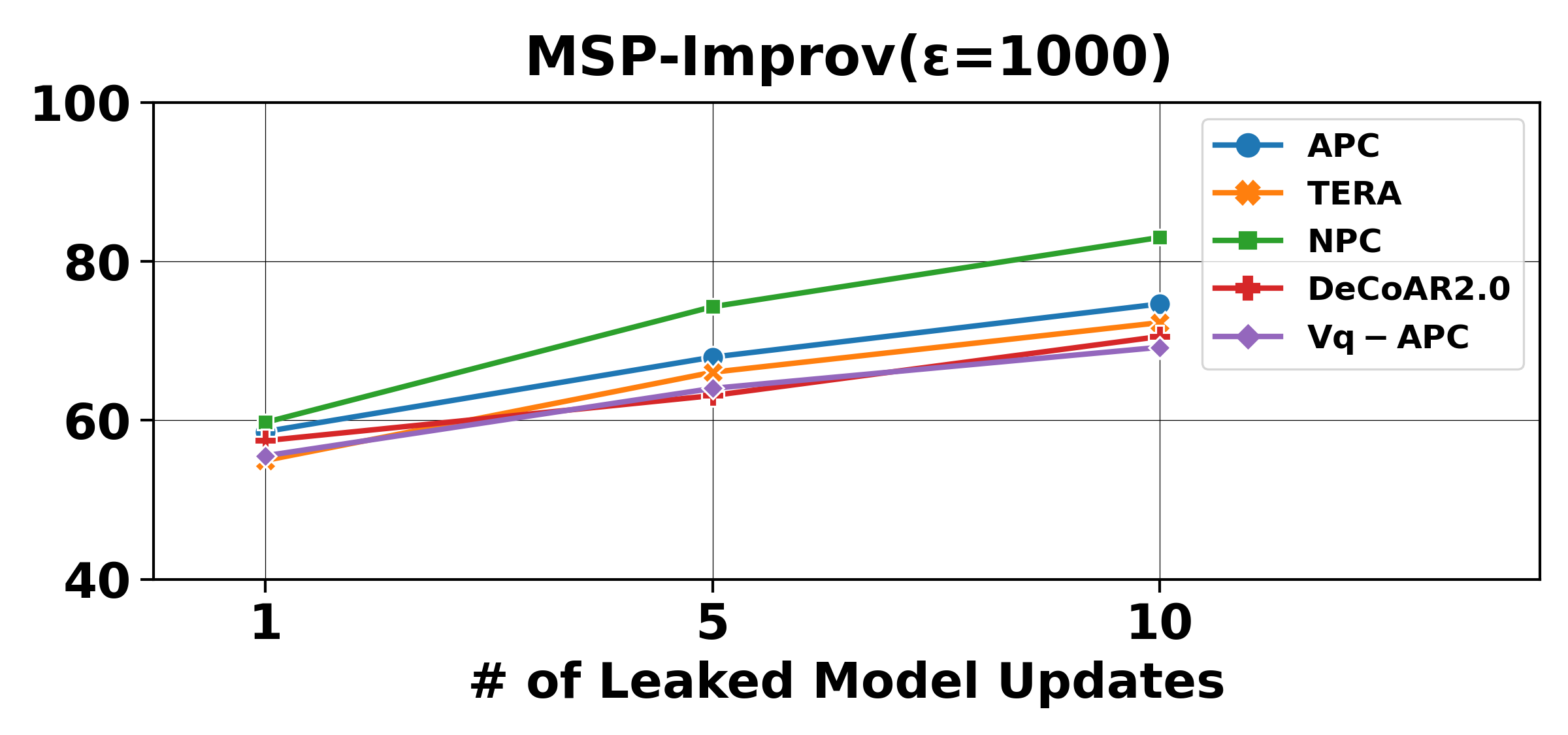}};
        
        \node[draw=none,fill=none] at (0, 5.2){\includegraphics[width=0.25\linewidth]{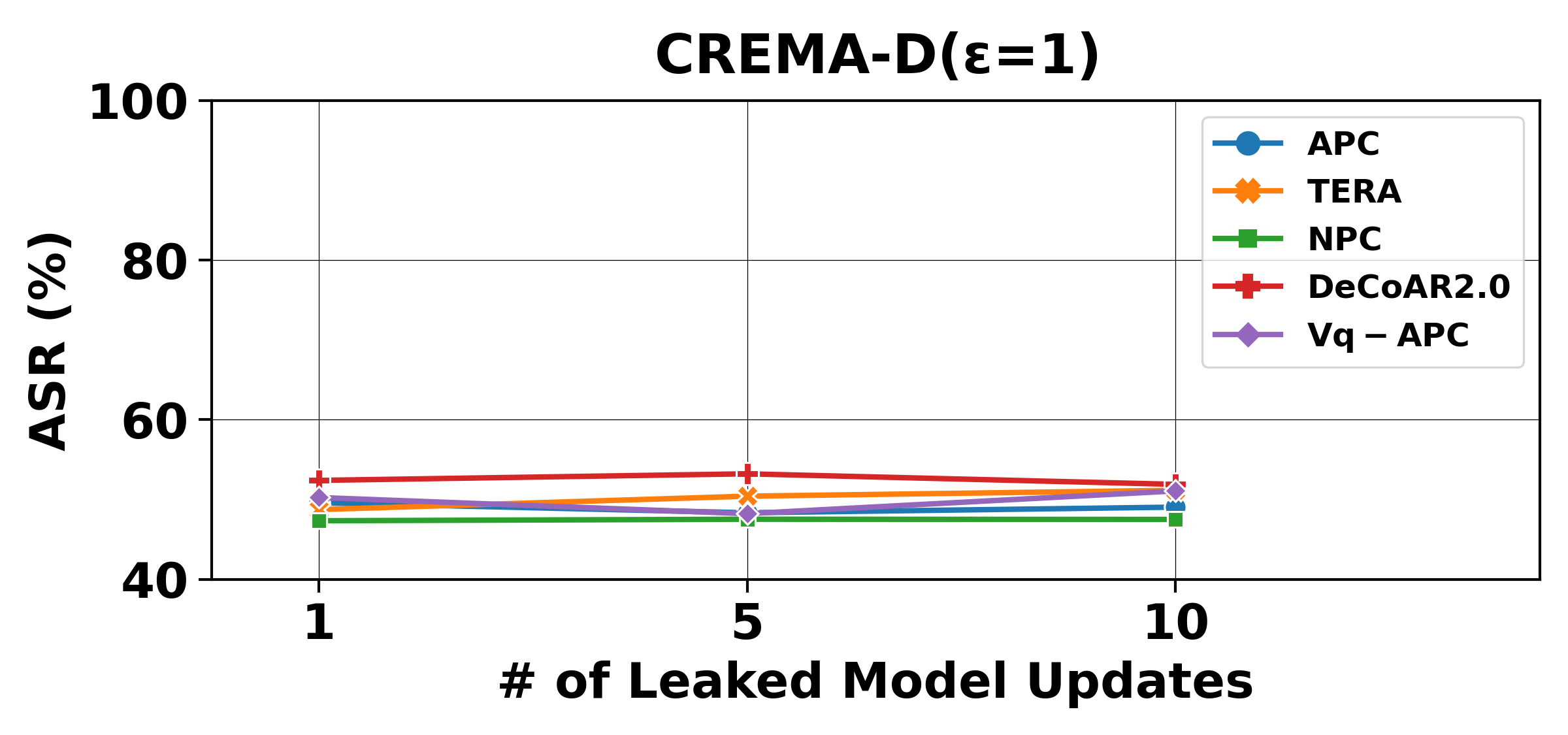}};
        
        \node[draw=none,fill=none] at (0.25\linewidth, 5.2){\includegraphics[width=0.25\linewidth]{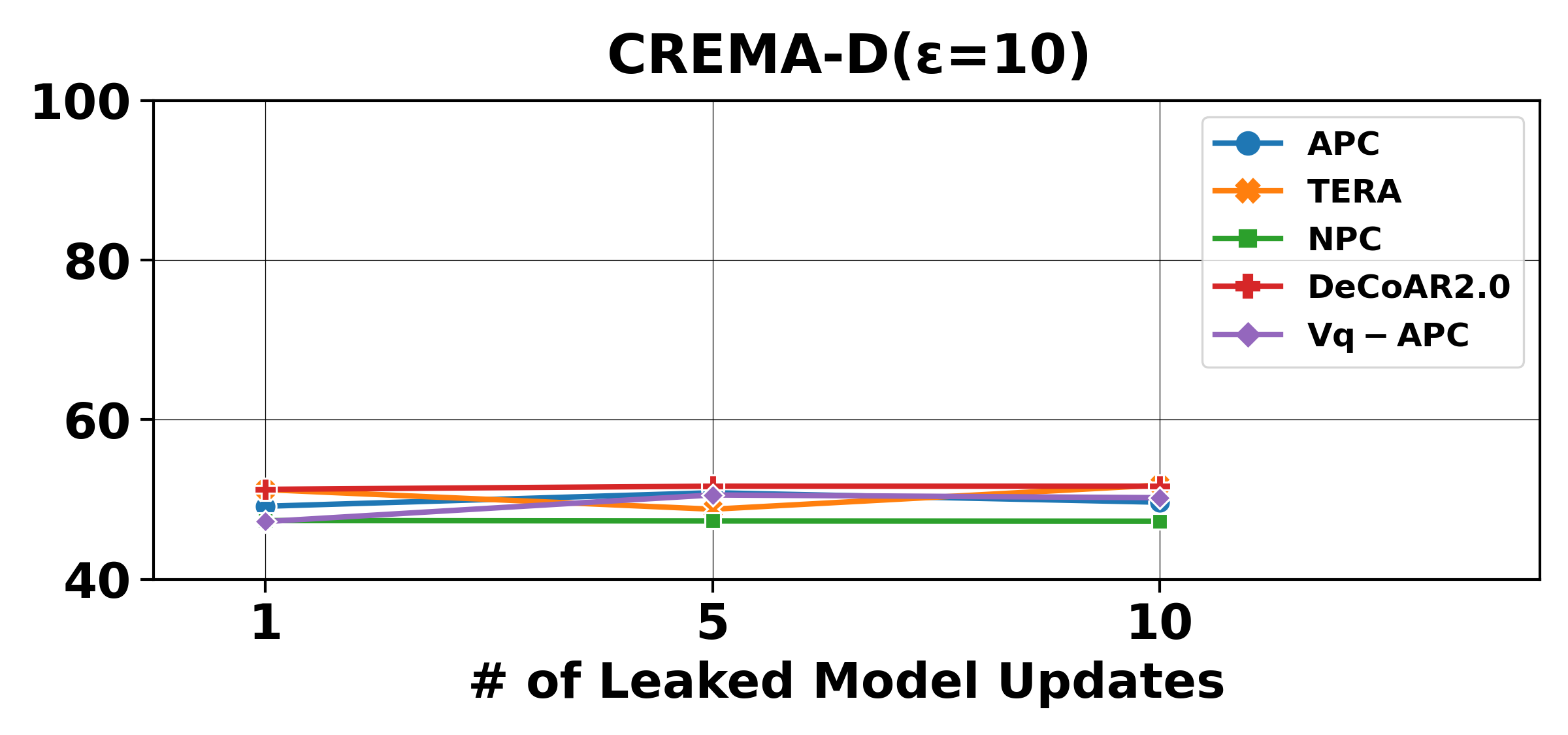}};

        \node[draw=none,fill=none] at (0.5\linewidth, 5.2){\includegraphics[width=0.25\linewidth]{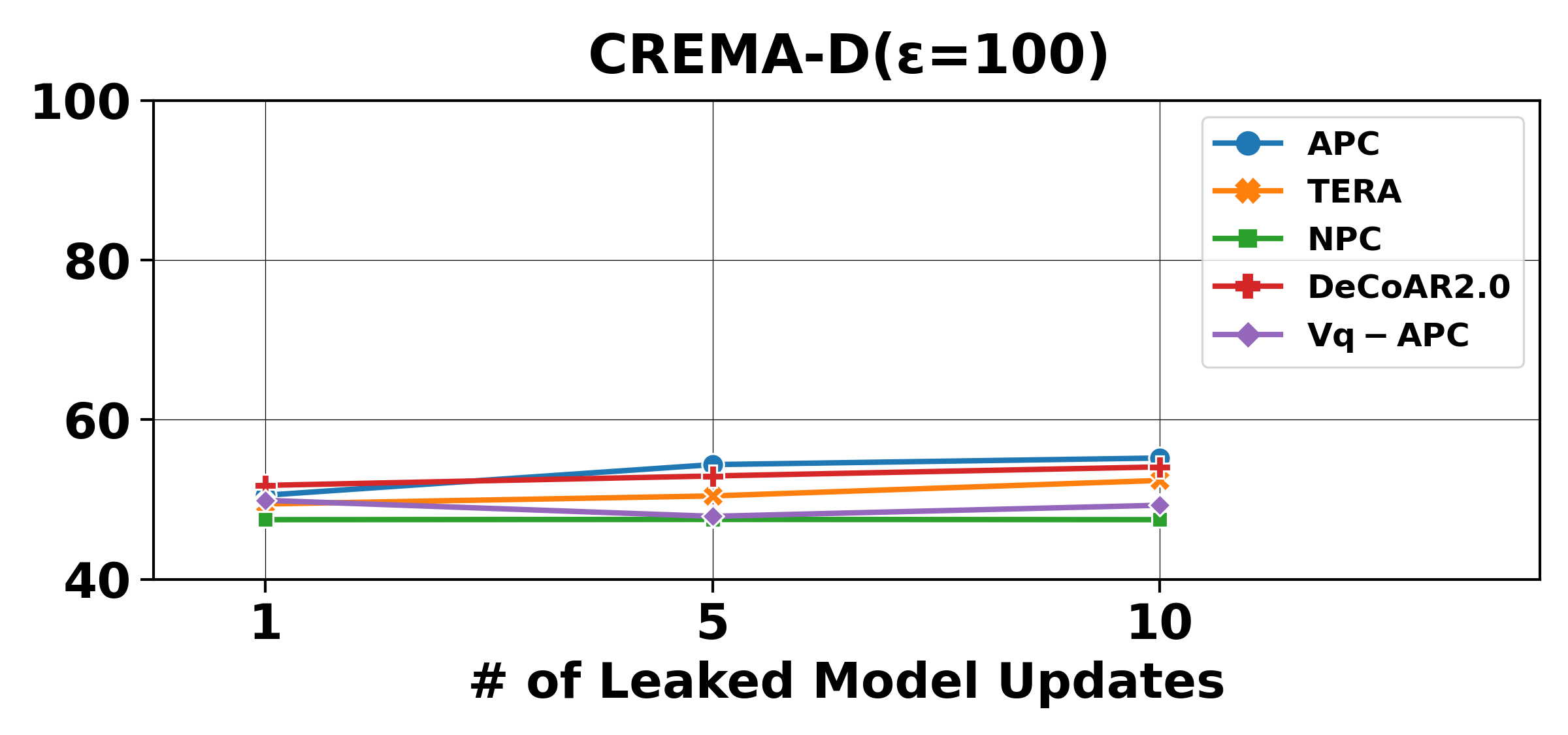}};

        \node[draw=none,fill=none] at (0.75\linewidth, 5.2){\includegraphics[width=0.25\linewidth]{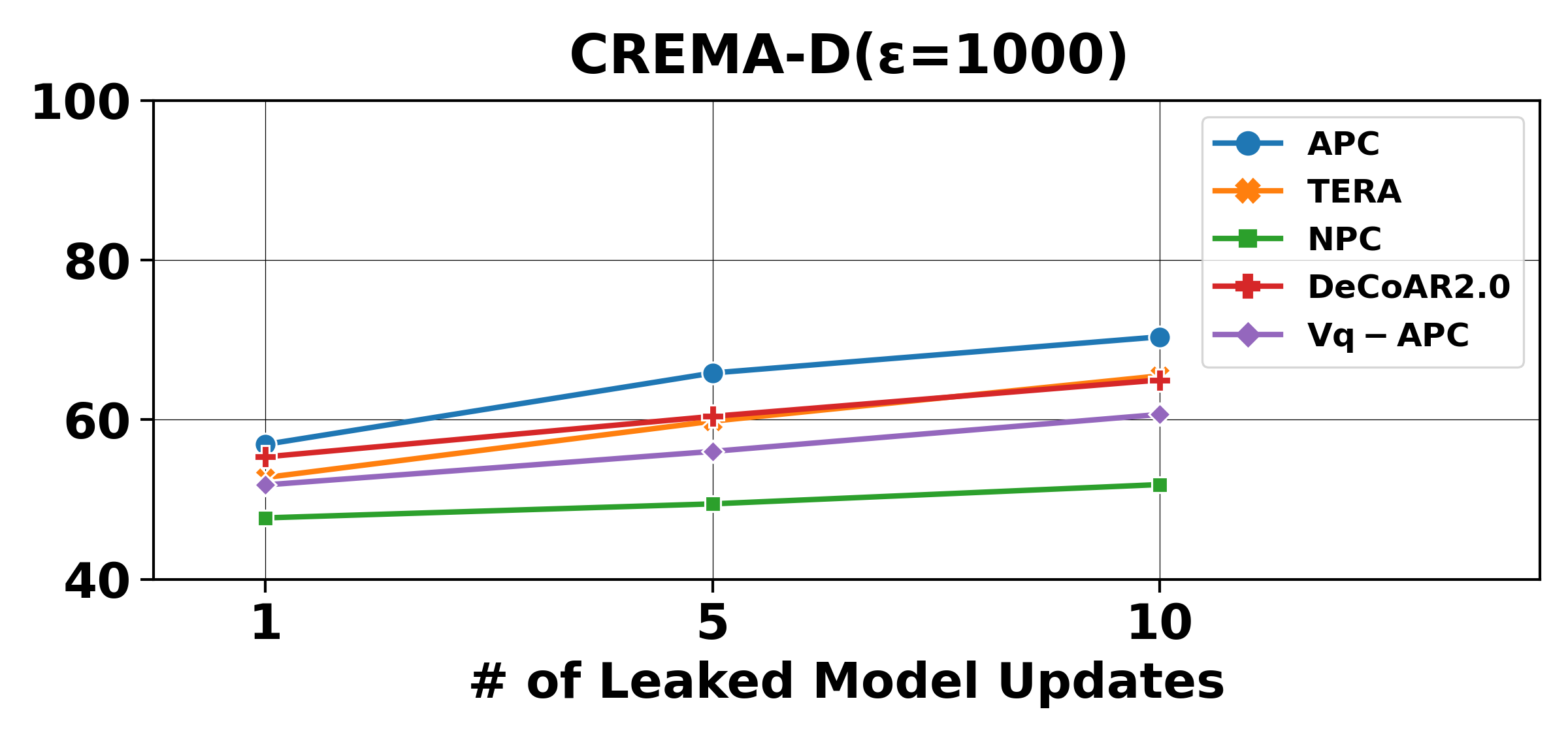}};
        
        \node[draw=none,fill=none] at (0, 7.5){\includegraphics[width=0.25\linewidth]{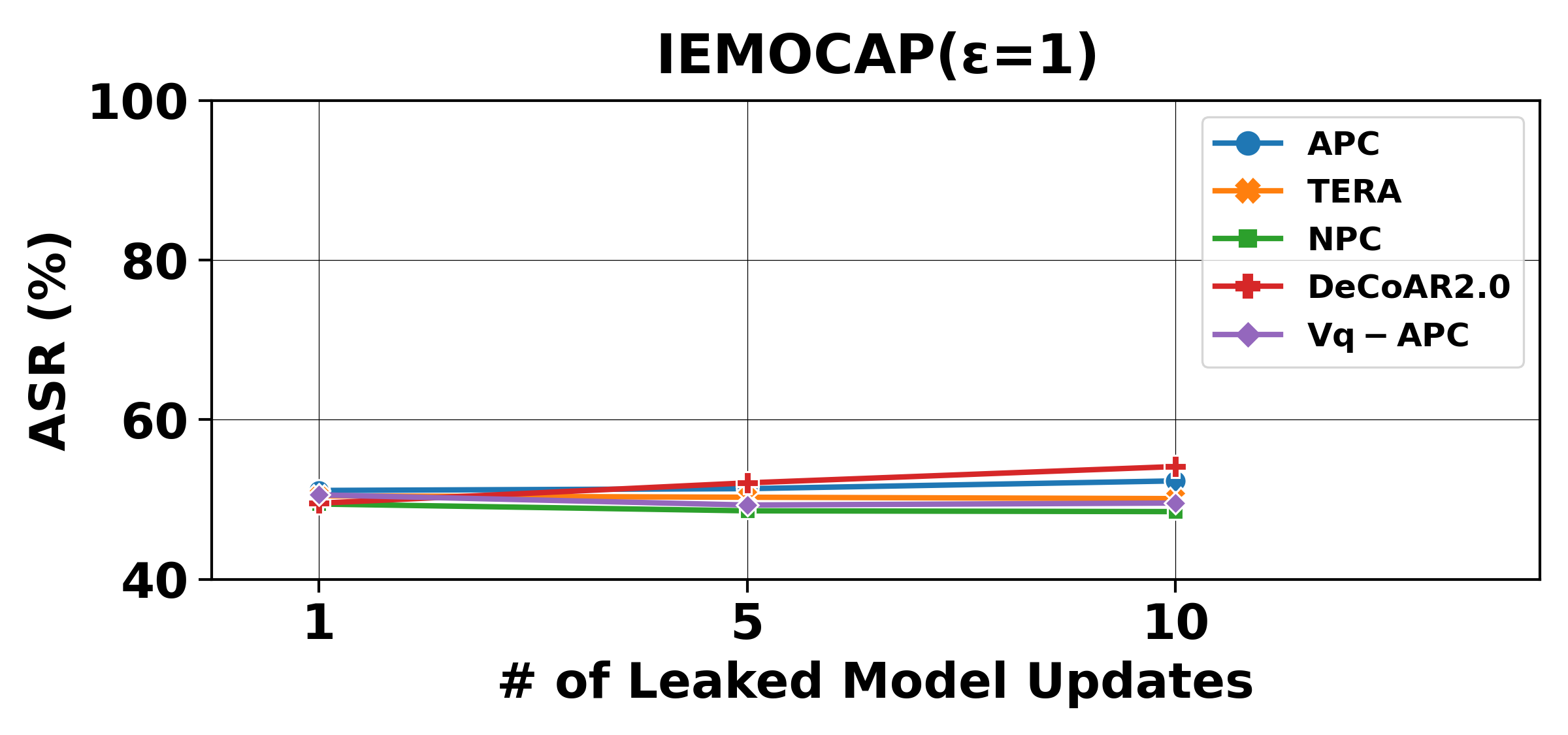}};
        
        \node[draw=none,fill=none] at (0.25\linewidth, 7.5){\includegraphics[width=0.25\linewidth]{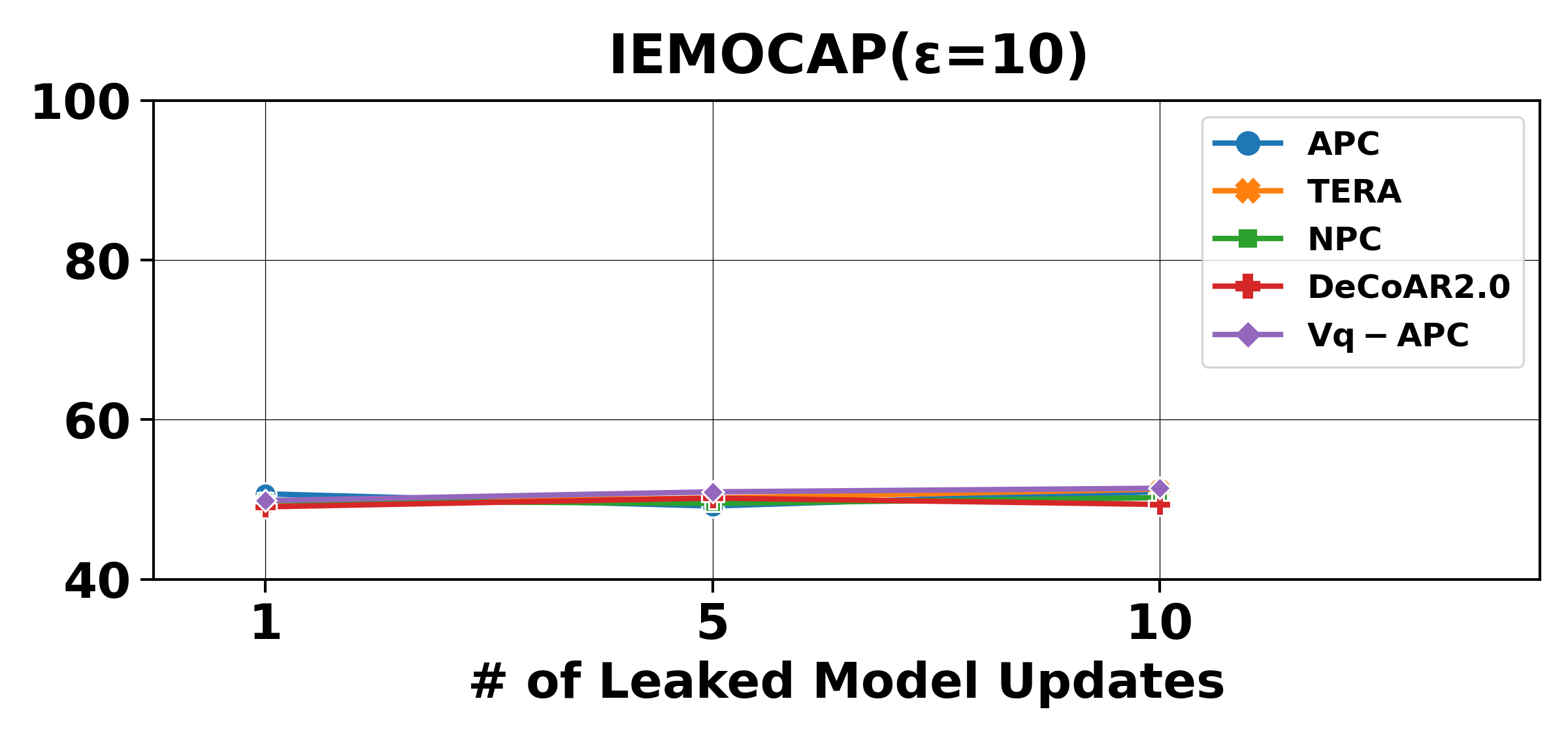}};

        \node[draw=none,fill=none] at (0.5\linewidth, 7.5){\includegraphics[width=0.25\linewidth]{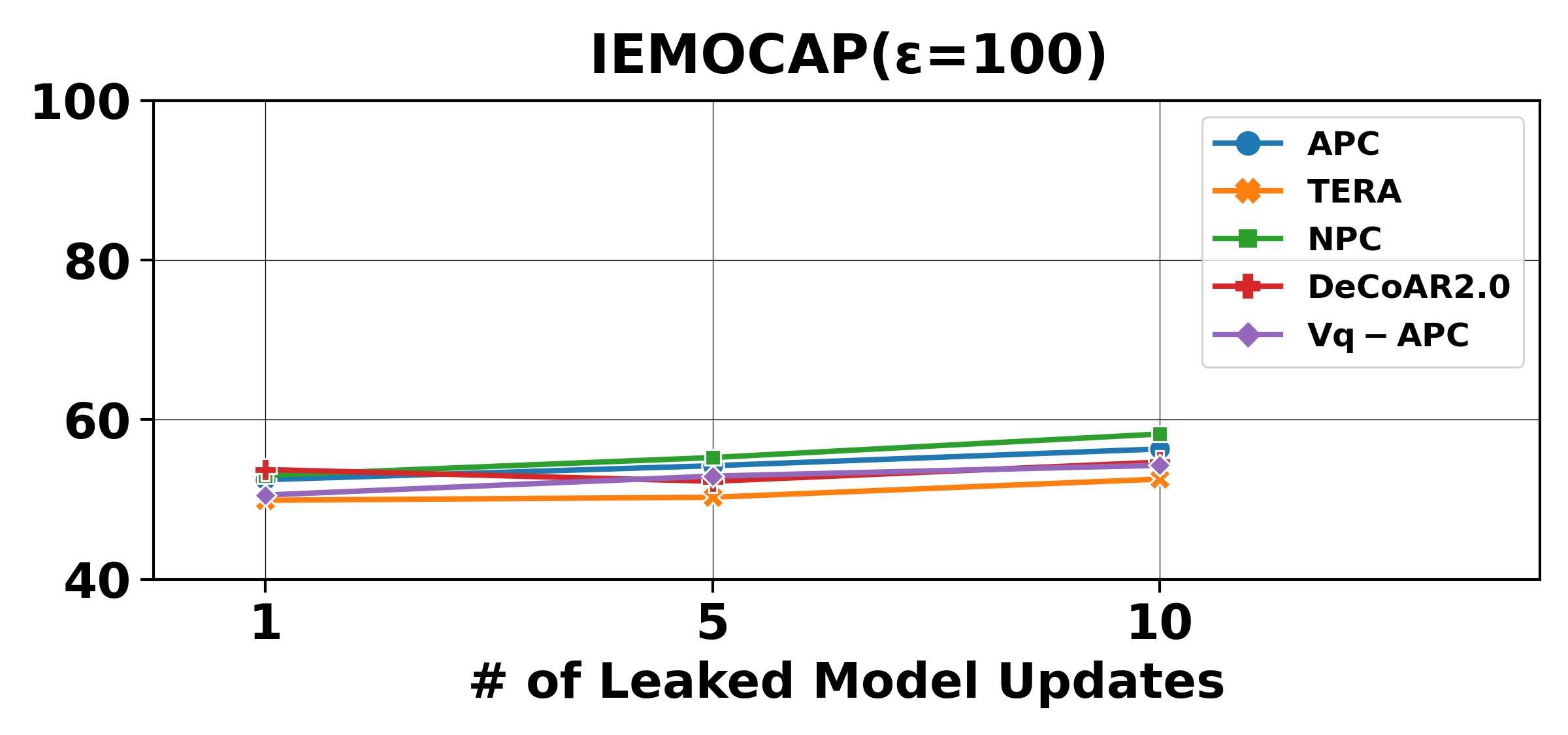}};

        \node[draw=none,fill=none] at (0.75\linewidth, 7.5){\includegraphics[width=0.25\linewidth]{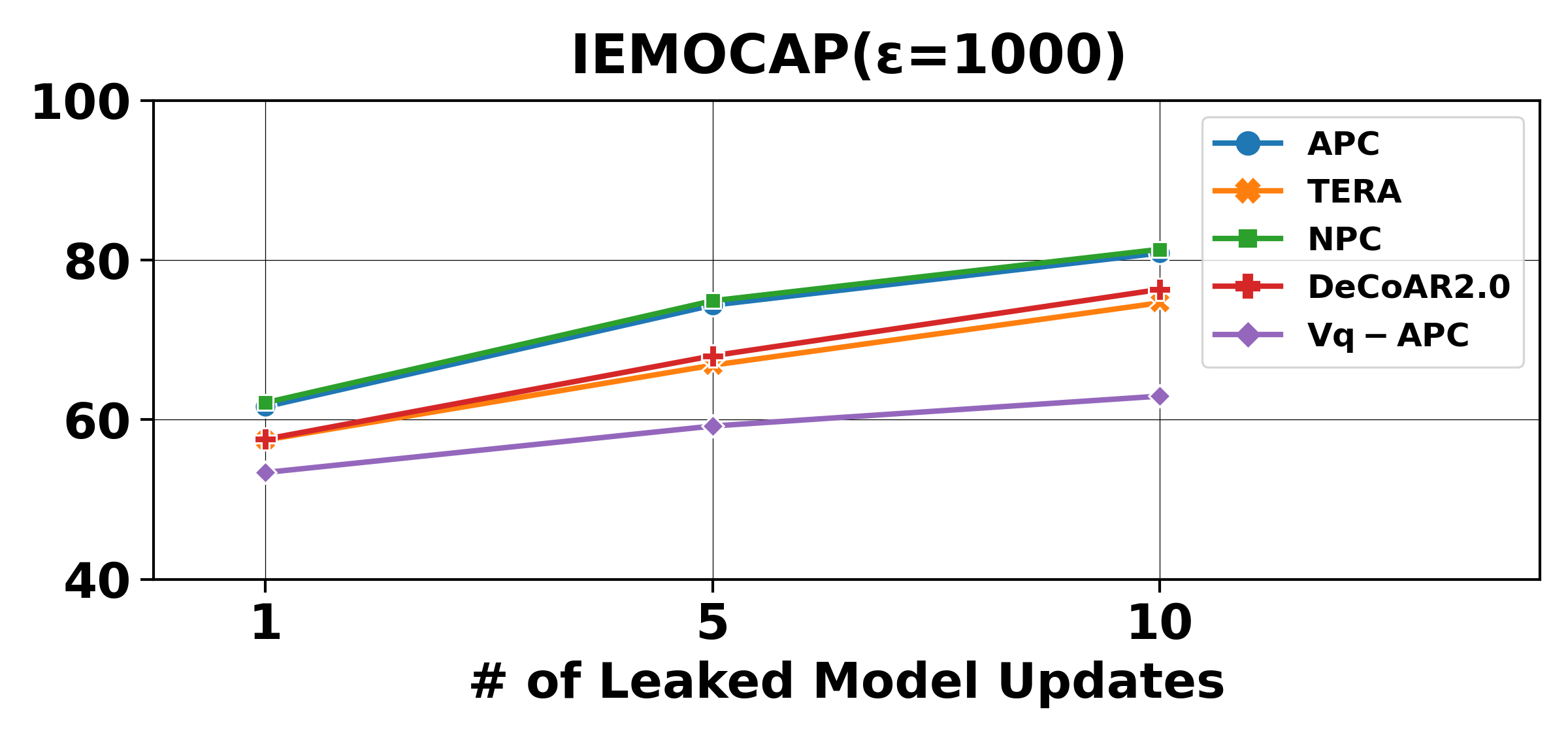}};
        
    \end{tikzpicture}
    
    \vspace{-2.5mm}
    \caption{The figure shows the attribute inference task at different privacy levels ($\epsilon\in\{1, 10, 100, 1000\}$) and with different numbers of leaked model updates. A larger $\epsilon$ indicates a weaker privacy guarantee.} 
    \label{fig:attack_dp_ablation}

    \vspace{-3.75mm}
    
} \end{figure*}

\noindent \textbf{Attacker Performance with access to multiple updates:} In the above mitigation, we explore the DP mitigation when the attacker has access to only one round of model updates. However, in our prior work \cite{feng2022user}, we have shown that the attacker can regain the ability to infer gender through aggregating multiple rounds of training updates with a weaker privacy guarantee. Similarly, we explore the ASR where the attacker can access multiple rounds of model updates in \autoref{fig:attack_dp_ablation}. We can find that the attacker can indeed regain the ability to infer gender with access to multiple rounds of training updates at a weaker privacy guarantee ($\epsilon=1000$). However, the attacker fails to infer gender when the privacy guarantee is strong. This validates that DP can effectively mitigate the proposed attribute inference attack when the attacker can only access one round of training updates. The DP mitigation becomes less effective with multiple rounds of training updates leaked to the attacker, where the attacker is able to perform the privacy attack under a weaker privacy guarantee. On the other hand, we notice that the performance drop in the SER application becomes substantial when $\epsilon\leq10$. These observations imply that DP can provide satisfied privacy protection against our proposed privacy attacks but with a noticeable drop in SER performance.

\section{Conclusion}

In this paper, we investigated attribute inference attacks on speech emotion recognition models trained within federated learning scenarios of shared gradient (FedSGD) and shared model (FedAvg). Our results show that unintended, and potentially private, properties (like gender) associated with the clients in the FL can leak through the shared updates when training the SER model. The deep models appear to internally capture many uncorrelated features with the tasks they are being trained for. Consequently, the attribute inference attacks are potentially powerful in this setting because the shared training updates carry significant potentially sensitive information about a (training) client. Our results suggest that the attacks are stronger in training the global SER model using the FedSGD algorithm than the FedAvg algorithm. We also show that the shared updates between the input and first dense layer leaks most information in this attribute inference attack. We further empirically demonstrate that defense strategies like dropout are not effective in mitigating this information leakage. We then show that Differential Privacy (DP) can mitigate this privacy attack with a stronger privacy budget by sacrificing the utility of the SER model.

These results motivate future work on defenses using the adversarial training technique to unlearn the sensitive attribute. Some of the limitations of our study include the relatively small number of clients and data sets even by combining three widely used SER test-beds. In addition, our work considers that attacker has access to each client's model updates, but this can be mitigated by aggregating shared updates from several clients in a local aggregator before transferring them to the central aggregator. In the future, we aim to build our SER model using more complex model structures, e.g.,  RNN+classifer. We also wish to apply the defense mechanism, such as adversarial training shown in \cite{jaiswal2020privacy}, to train the SER model in the FL setup. Meanwhile, the current attack model utilizes only two public SER data sets, and we aim to include more public data sets to further increase the attacker's performance. Finally, we wish to evaluate the membership inference attack \cite{hu2021source} and label inference attack \cite{wainakh2021user} within similar experimental settings.

\section{Acknowledgment}
\label{sec:ack}
The work was supported by the USC-Amazon Center on Trusted AI.

% or


% \appendices

% \section{}

% \ifCLASSOPTIONcompsoc
  %\section*{Acknowledgments}
%\else
  %\section*{Acknowledgment}
% \fi


\ifCLASSOPTIONcaptionsoff
  \newpage
\fi



\bibliographystyle{IEEEtran}
\bibliography{refs.bib}

% begin{IEEEbiography}[{\includegraphics[width=1in,height=1.25in,clip,keepaspectratio]{mshell}}]{Michael Shell}

\vspace{-1mm}

\begin{IEEEbiography}[{\includegraphics[width=1.25in,height=1.25in,clip,keepaspectratio]{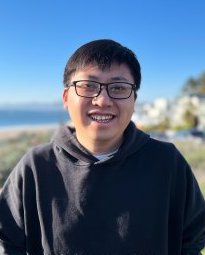}}]{Tiantian Feng}
received the B.S. degree in instrument technology
from the Nanjing University of Posts and Telecommunications, Nanjing, China,
in 2013 and the M.S. degree, in 2015 in electrical engineering from the University of Southern California, Los Angeles, CA, USA, where he is currently working toward the Ph.D. degree. His current research interests include privacy-enhancing computing, smart wearable sensing applications, multimodal biomedical signal processing, and affective computing.
\end{IEEEbiography}

\vspace{-10mm}

\begin{IEEEbiography}[{\includegraphics[width=1.05in,height=1.5in,clip,keepaspectratio]{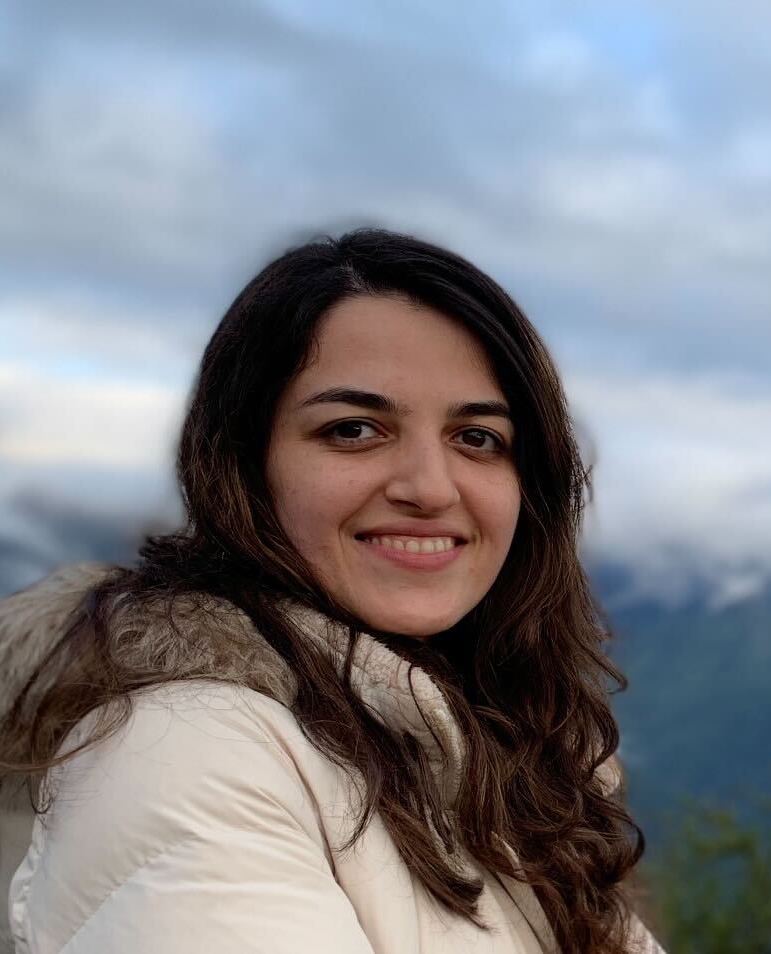}}]{Hanieh Hashemi}

is a Ph.D Candidate in the ECE department at the University of Southern California. She is advised by Professor Murali Annavaram. Her thesis focuses on Data Privacy in Deep Neural Networks and Recommendation Systems. She worked with Facebook AI Research on recommendation system privacy. She collaborated with Samsung Semiconductor on efficient graph processing methods.

\end{IEEEbiography}

\vspace{-5mm}
\begin{IEEEbiographynophoto}{Rajat Hebbar}
received the bachelor's degree in electronics and communication engineering from the National Institute of Technology, Karnataka, India, and the master's degree in electrical and computer engineering from the University of Southern California, Los Angeles, CA, USA, where he is currently working toward the Ph.D. degree with Electrical and Computer Engineering Department. His research interests include developing machine learning-based robust speech and audio processing techniques for several challenging real-world domains, such as multimedia, wearable-device audio, and multiparty meetings.
\end{IEEEbiographynophoto}

% \begin{IEEEbiography}[{\includegraphics[width=1.3in,height=1.3in,clip,keepaspectratio]{Photo/rahul_photo.png}}]{Rahul Gupta}

% is a Senior Applied Science manager at the Spoken Language Understanding Innovations (SLU-Innovations) team in Cambridge, Massachusetts. Since joining the Alexa organization, he has focused on designing NLU models for scalability and speed. Some of his more recent research focuses on Trustworthy Machine Learning with emphasis on privacy preserving techniques, fairness and federated learning. He received his PhD from the University of Southern California in 2016 on interpreting non-verbal communications in human interaction. He has published several papers at avenues such as EMNLP, ACL, NAACL, ACM Facct and IEEE-Transactions. He is also co-inventor on ten patent pending technologies at Amazon.

% \end{IEEEbiography}

\vspace{-10mm}
\begin{IEEEbiographynophoto}{Murali Annavaram}
is the Dean's Professor at the University of Southern California, Electrical and Computer Engineering and Computer Science departments. He currently also holds the Rukmini Gopalakrishnachar Visiting Chair Professor at the Indian Institute of Science.  His research focuses energy efficiency through heterogeneous computing, near-data computing, hardware-assisted secure and private machine learning, and superconducting electronics design. He is the founding Director of the USC-Meta Center for Research and Education in AI and Learning. He is also the architecture thrust leader for the DISCoVER NSF Expeditions in Computing Center at USC. For his numerous publications, he was inducted into the hall of fame at three top-tier computer architecture conference venues, ISCA, HPCA and MICRO. More information about his work can be found on https://annavar.am/.
\end{IEEEbiographynophoto}

\vspace{-10mm}
\begin{IEEEbiography}[{\includegraphics[width=1.28in,height=1.28in,clip,keepaspectratio]{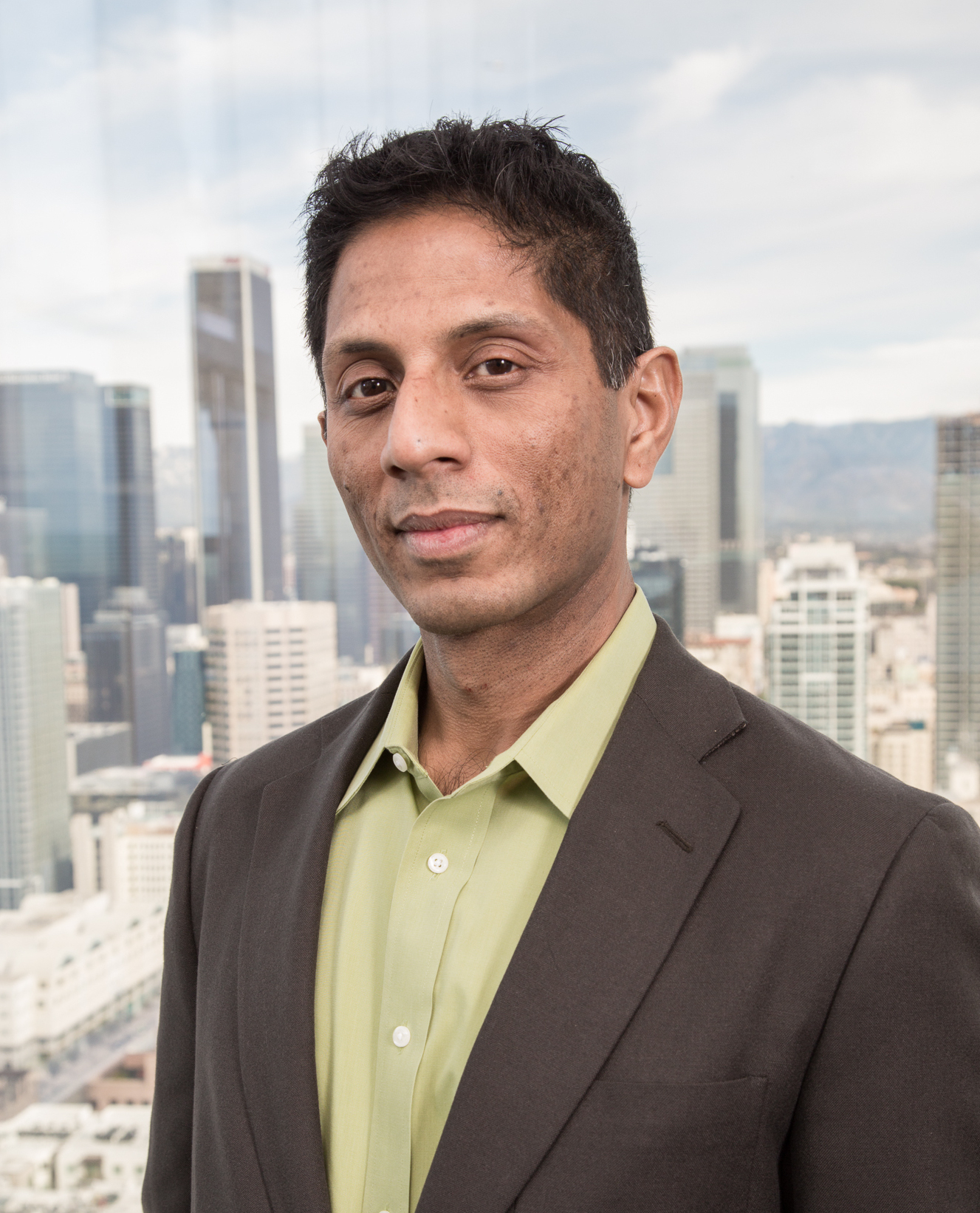}}]{Shrikanth (Shri) Narayanan}
(StM'88–M'95–SM'02–F'09) is the University Professor and Niki \& C. L. Max Nikias Chair in Engineering at the University of Southern California, and holds appointments as Professor of Electrical and Computer Engineering, Computer Science, Linguistics, Psychology, Neuroscience, Otolaryngology-Head and Neck Surgery, and Pediatrics, Research Director of the Information Science Institute, and director of the Ming Hsieh Institute. Prior to USC he was with AT\&T Bell Labs and AT\&T Research. Shri Narayanan and his Signal Analysis and Interpretation Laboratory (SAIL) at USC focus on developing engineering approaches to understand the human condition (including voice, face and biosignal based biometrics) and in creating machine intelligence technologies that can support and enhance human experiences. He is a Guggenheim Fellow and a Fellow of the Acoustical Society of America, IEEE, ISCA, the American Association for the Advancement of Science (AAAS), the Association for Psychological Science, the American Institute for Medical and Biological Engineering (AIMBE) and the National Academy of Inventors. He has published over 900 papers and has been granted eighteen U.S. patents.
\end{IEEEbiography}

%\newpage



%\vfill


\end{document}